\documentclass[preprint]{aastex}
\usepackage{mkfig}
\begin{document}

\title{Microstructure of the Local Interstellar Cloud and the Identification 
of the Hyades Cloud.}

\author{Seth Redfield}
\affil{JILA, University of Colorado}
\affil{Boulder, CO 80309}
\email{sredfiel@casa.colorado.edu}
\author{Jeffrey L. Linsky}
\affil{JILA, University of Colorado and NIST}
\affil{Boulder, CO 80309}
\email{jlinsky@jila.colorado.edu}

\begin{abstract}

We analyze high-resolution UV spectra of the Mg~II h and k lines for 18 
members of the Hyades Cluster to study inhomogeneity along these proximate 
lines of sight.  The observations were taken by the Space Telescope Imaging 
Spectrograph (STIS) instrument on board the Hubble Space Telescope (HST).  
Three distinct velocity components are observed.  All 18 lines of sight show 
absorption by the Local Interstellar Cloud (LIC), ten stars show absorption 
by an additional cloud, which we name the Hyades Cloud, and one star exhibits 
a third absorption component.  The LIC absorption is observed at a lower 
radial velocity than predicted by the LIC velocity vector derived by 
\citet{lall92} and \citet{lall95}, ($v_{\rm predicted}(\rm LIC) - 
v_{\rm observed}(\rm LIC) = 2.9~\pm~0.7$~km~s$^{-1}$), which may indicate a 
compression or deceleration at the leading edge of the LIC.  We propose an 
extention of the Hyades Cloud boundary based on previous HST observations of 
other stars in the general vicinity of the Hyades, as well as ground-based 
Ca~II observations.  We present our fits of the interstellar parameters for 
each absorption component.  The availability of 18 similar lines of sight 
provides an excellent opportunity to study the inhomogeneity of the warm, 
partially ionized local interstellar medium (LISM).  We find that these 
structures are roughly homogeneous.  The measured Mg~II column densities do 
not vary by more than a factor of 2 for angular separations of 
$\lesssim~8^{\circ}$, which at the outer edge of the LIC correspond to 
physical separations of $\lesssim~0.6$~pc. 

\end{abstract}


\section{Introduction}

The morphology and physical properties of our Local Interstellar Medium 
(LISM) are being actively studied but are still not fully understood.  Many 
recent high-resolution observations have uncovered hints of a complex LISM 
structure that have yet to be synthesized into a coherent picture.  That the 
structure of the LISM, extending $\sim~250$~pc from the Sun, is not a 
homogeneous medium has been demonstrated by \citet{fy91} and by \citet{djp95},
 from comparing measurements of H~I column density ($N_{\rm HI}$) with 
distance.  In comparisons of Na~I column density observations, \citet{sf99} 
demonstrated that there is very little cold neutral material within 50~pc of 
the Sun.  They interpreted the rapid accumulation of Na~I column density at 
distances $>~50$~pc as indicating the boundary of the Local Bubble, a 
superbubble cavity of hot ionized material produced by the OB star winds and 
supernovae of the Scorpius-Centaurus association \citep{frisch95,lb96}.  The 
inhomogeneous variation of H~I column density measured within the Local 
Bubble indicates that warm, partially ionized material exists within the 
Local Bubble as a complex of cloudlet structures.  

The simplest interpretation of the structure of the complex of warm, 
partially ionized material is that of individual clouds, which in this simple 
model would be coherent in density, velocity, temperature, and abundances.  
\citet{lall92} and \citet{lall95} found a coherent velocity structure for the 
warm gas cloud that directly surrounds the solar system.  They called this 
cloud the Local Interstellar Cloud (LIC).  The LIC velocity vector agrees 
with the flow of material through the heliosphere, and into the solar system, 
as measured by \citet{witte93}, implying that the Sun is located inside the 
LIC.  The LIC velocity vector has also been successful in predicting central 
absorption velocities for many lines of sight through the LIC.  The 
temperature structures of different clouds in the LISM are studied by 
comparing the Doppler width of absorption by atoms and ions of different mass.
  Absorption for LIC lines of sight indicates a temperature of 
8000~$\pm$~1000~K \citep{lin93,lall94,lin95,gry95,pisk97,wood98}.  

\citet{lin00} summarized the hydrogen and deuterium column densities through 
the LIC along many lines of sight toward nearby stars inferred from spectra 
obtained with the Goddard High Resolution Spectrograph (GHRS) and Space 
Telescope Imaging Spectrograph (STIS) instruments on the Hubble Space 
Telescope (HST) and the Extreme Ultraviolet Explorer (EUVE).  Taking advantage
 of the constant D/H ratio in the LIC, and assuming a constant H~I density in 
the cloud, \citet{lin00} estimated the distance to the edge of the LIC for 
many lines of sight.  \citet{red00} synthesized the results and were 
successful in fitting a quasi-spheroid structure to the observed data.  This 
model has successfully predicted hydrogen column densities along various lines
 of sight \citep{woodkdwarf}.  

The first-order assumptions involved in the cloudlet model of the LISM may be 
too simplistic.  Using ultra-high resolution Ca~II spectra of nearby stars, 
\citet{welsh98} presented preliminary results that indicate a complex 
velocity structure of inhomogeneous absorption.  A highly inhomogeneous, 
filamentary structure has already been proposed to explain abundance anomalies
 and the LISM velocity structure \citep{frisch81,frisch96,frisch00}.  
Shock-front destruction of dust grains can in principle account for the 
observed abundance anomalies, and the general LISM velocity structure appears 
to be consistent with the outflow of material from the Scorpius-Centaurus 
association.  

Measurements of the degree of inhomogeneity in the LISM are needed to 
understand the structure and morphology of the LISM.  One test of homogeneity 
involves a comparison of interstellar medium (ISM) parameters along similar 
lines of sight.  For example, \citet{frail94} studied observations of 
absorption toward high-velocity pulsars in the 21-cm H~I fine structure line. 
 The high transverse velocity of these objects actually shifts the line of 
sight by tens of AU~yr$^{-1}$.  Changes in the absorption spectra indicate 
small-scale structure on scales of 5-100~AU.  Most of these pulsars are 
located well outside of the Local Bubble (d~$=~55-2650$~pc), so these 
inhomogeneities are most likely not associated with nearby ($<~50$~pc) cloud 
material.  

Another technique for studying similar lines of sight is to observe wide 
binary or multiple star systems.  \citet{watson96} and \citet{meyer96} have 
observed Na~I and Ca~II absorption variations for multiple systems with 
stellar separations of 480 to 29,000~AU.  Observations of other ions dominant 
in the cold neutral medium (Cr~II and Zn~II) by \citet{lauro98} show no 
interstellar absorption variation, making it difficult to ascertain the 
nature of the variable material.  However, most of these binaries are located 
more than 100~pc away, and may be at the edge or outside of the Local Bubble. 
 The observed small scale structure is therefore likely produced in the cold 
neutral medium component of the ISM, near the boundary of the Local Bubble or 
beyond.  Therefore, there is no evidence for significant variations in the 
warm, partially ionized material in the solar neighborhood along similar sight
 lines.  Likewise, the nearest multiple star system ($\alpha$~Cen~A, 
$\alpha$~Cen~B, and Proxima Cen), shows no sign of variations in interstellar 
absorption over time, or among the similar sight lines of each star 
\citep{lin96,wood01}.  At a mean distance of 1.3~pc, the physical separations 
of the two $\alpha$~Cen stars is only 24~AU, while Proxima~Cen is 12,000~AU 
away from the other two stars.  

Although the inhomogeneity of the LISM is central to our understanding of the 
ISM, most observations to date (e.g., Na~I and Ca~II) have sampled primarily 
the cold neutral component near the boundary of the Local Bubble.  There has 
yet to be a significant test of the homogeneity of the warm, partially ionized
 LISM that directly surrounds the solar system.  In this paper we present 
observations of a sample of 18 stars in the Hyades cluster.  As members of the
 same stellar cluster, they share similar sight lines.  This data set will 
directly address the inhomogeneity of the LIC and other warm material in the 
LISM, and provide a test of our current understanding of the morphology and 
physical properties of the LISM.  The STIS spectra are described in \S~2, and 
we present in \S~3 our determination of the interstellar parameters.  In \S~4,
 we discuss the implications of the measured interstellar parameters for the 
inhomogeneity of the LIC.

\section{Observations}

The Space Telescope Imaging Spectrograph (STIS) instrument aboard HST is 
described by \citet{kimble98} and \citet{woodgate98}.  STIS observed 18 F 
dwarf members of the nearby Hyades star cluster.  Details regarding the 
observations are given in Table 1.  This dataset was acquired under observing 
program 7389 with principal investigator E. Bohm-Vitense.  All exposures were 
taken with the E230H grating and the 0.1'' x 0.2'' or 0.2'' x 0.2'' aperture 
to attain a spectral resolution of 2.6 $\rm km~s^{-1}$.  Within the 2674-2945 
\AA\ observed spectral range, only the Mg~II h and k lines show identifiable 
interstellar absorption features.  The S/N ratios near the peak of these lines
 range from 5-20.

We reduced the data acquired from the HST Data Archive using the STIS team's 
CALSTIS software package written in IDL \citep{lindler99}.  The reduction 
included assignment of wavelengths using calibration spectra obtained during 
the course of the observations.  The ECHELLE\_SCAT routine in the CALSTIS 
software package was used to remove scattered light.  However, the scattered 
light contribution is negligible in this spectral range, and does not 
influence the uncertainties in our spectral analysis. 

\section{Discussion}

\subsection{Spectral Analysis}	

Figures 1a, 1b, and 1c show the Mg~II h and k lines observed from the 18 
Hyades stars.  The vacuum rest wavelengths of the h and k lines are 
2803.531~\AA\ and 2796.352~\AA, respectively.  In Figures 1a, 1b, and 1c, the 
Mg~II lines are shown with a heliocentric velocity scale, together with the 
best fits to the absorption by each interstellar component (dotted lines), and
 the total interstellar absorption convolved with the instrumental profile 
(thick solid lines).  Although interstellar absorption is obvious for all of 
the lines of sight, sometimes two, and in one case (HD~28736), three velocity 
components are clearly visible.  

\begin{figure}
\plotfiddle{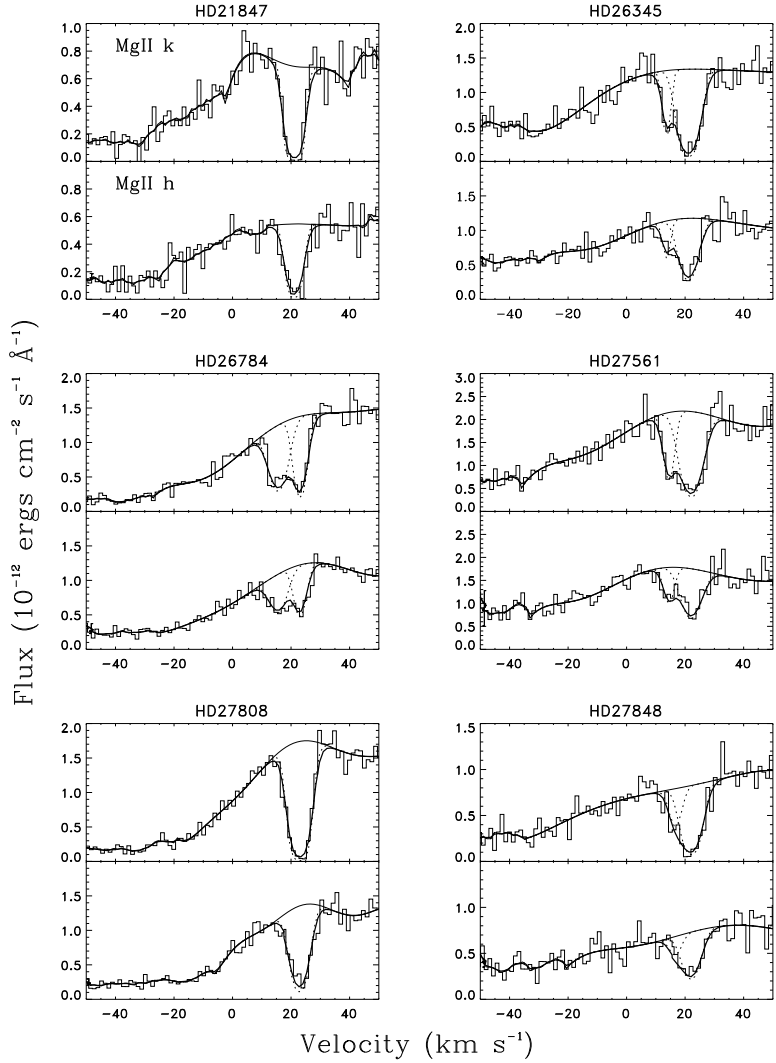}{5.in}{0}{55}{55}{-150}{10}
\figurenum{1a}
\figcaption[fig1a.eps]{Our best fits to the Mg~II h and k lines for six of 
the Hyades stars.  For each star, the Mg~II~k line is shown in the top panel, 
and the Mg~II~h line in the bottom panel.  The data are shown in histogram 
form.  The thin solid lines are our estimates for the missing stellar 
continuum across the absorption lines, computed by using polynomial fits to 
spectral regions blueward and redward of the absorption features.  The dotted 
lines are the best-fit individual absorption lines before convolution with the
 instrumental profile.  The thick solid line represents the combined 
absorption fit after instrumental broadening.  The spectra are plotted against
 heliocentric velocity.  Notice that for all these stars, the stellar Mg~II h 
and k emission lines are centered $\sim$~40 km~s$^{-1}$ to the red, due to 
their high radial velocities (see Table~2).  The absorption features at radial
 velocites ranging from 10-25 km~s$^{-1}$ therefore lie in the blue wing of 
the Mg~II lines.  The parameters involved in these fits are given in Table~3. 
\label{fig1a}}
\end{figure}

\begin{figure}
\figurenum{1b}
\plotfiddle{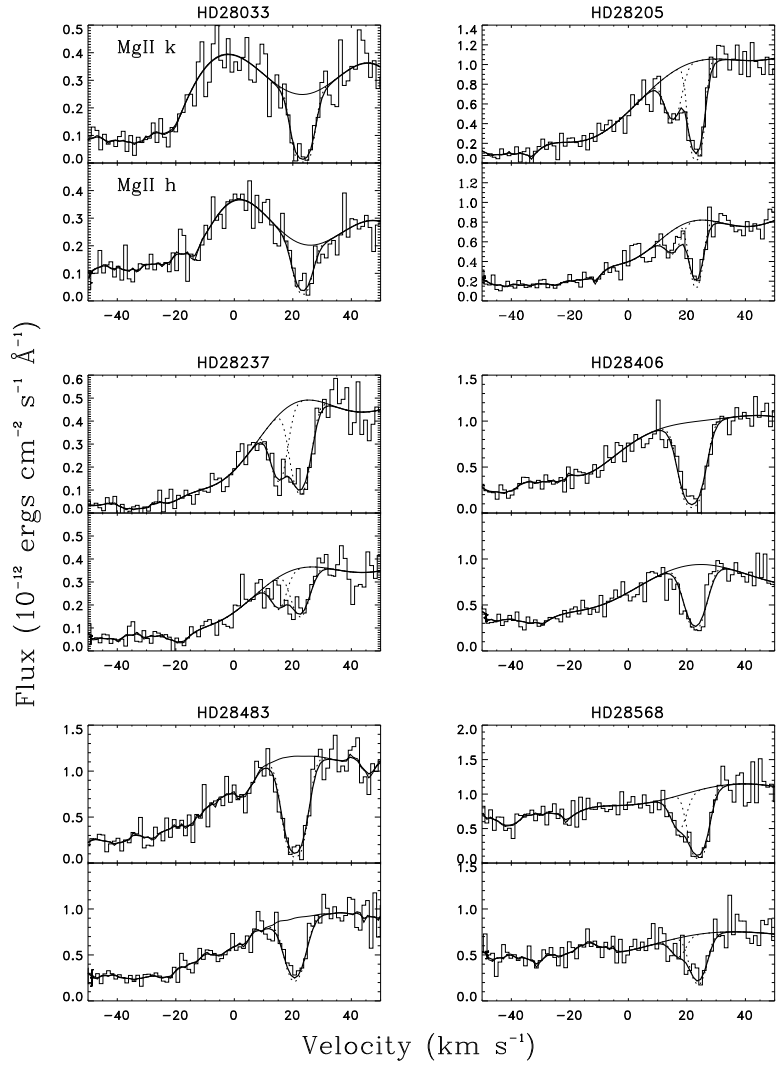}{5.in}{0}{55}{55}{-150}{10}
\figcaption[fig1b.eps]{As in Figure~\ref{fig1a}, but for six additional 
Hyades stars. \label{fig1b}}
\end{figure}

\begin{figure}
\figurenum{1c}
\plotfiddle{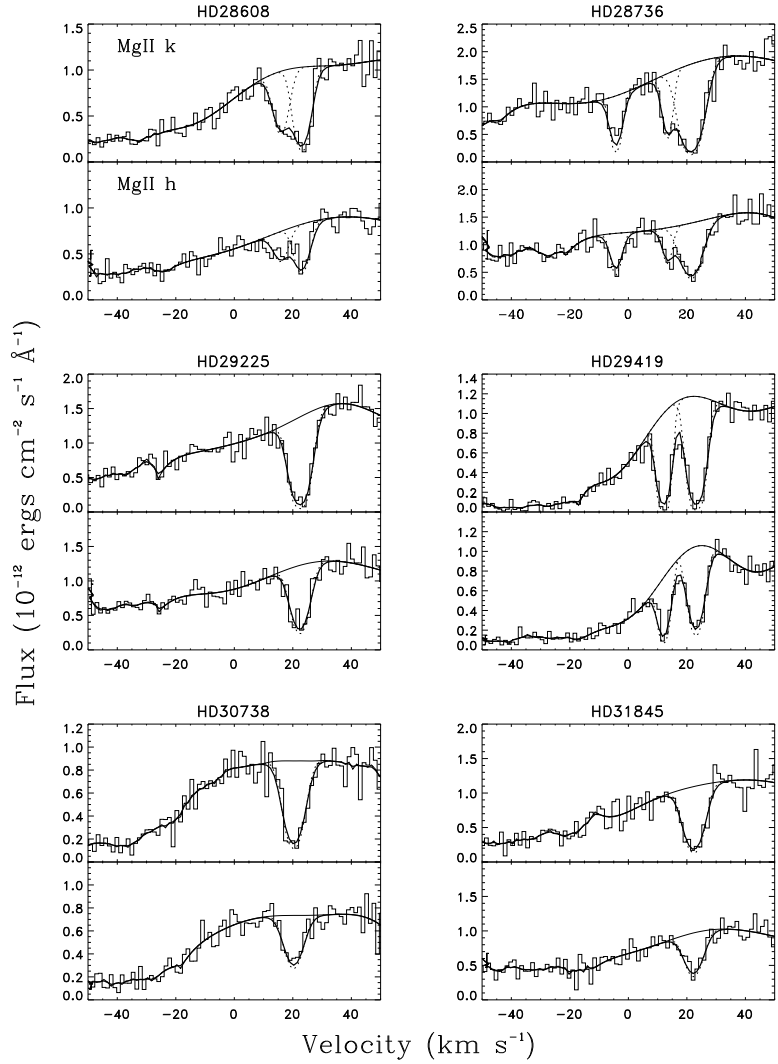}{5.in}{0}{55}{55}{-150}{10}
\figcaption[fig1c.eps]{As in Figure~\ref{fig1a}, but for the final six Hyades 
stars.  Note that the HD~28736 profile requires three absorption components. 
\label{fig1c}}
\end{figure}

\placefigure{fig1a}
\placefigure{fig1b}
\placefigure{fig1c}

All of the Hyades target stars have radial velocities $>$~30~km~s$^{-1}$.  
Therefore, the peaks of the Mg~II h and k emission lines are shifted 
$>$~30~km~s$^{-1}$ to the red.  Since all interstellar absorption occurs at 
velocities $<$~30~km~s$^{-1}$, the interstellar features appear on the blue 
side of the Mg~II lines.  This convenient placement of the interstellar 
absorption allows us to implement a straightforward method for estimating the 
missing stellar continuum across the absorption.  We fit a polynomial to 
spectral regions just blueward and redward of the interstellar absorption.  
Our estimated stellar emission is shown by the thin solid lines in Figures 1a,
 1b, and 1c.  

Once the stellar Mg~II emission line profiles have been estimated, we fit the 
interstellar absorption using standard techniques (cf. \citet{lin96,pisk97,
dring97}).  We start with only one absorption component in the fit, and 
increase the number of components as the data warrant, and the quality of the 
fit improves.  For 10 of the 18 Hyades stars adding a second component 
improved the $\chi^2$ metric significantly, while the spectral fit of only one
 star (HD~28736) improved significantly with the addition of a third 
component.  While, additional weak absorption components could be present, 
they do not appear in the data much above the level of noise, and thereby do 
not significantly influence the $\chi^2$ metric.  Therefore, we fit the 
spectra with the lowest number of absorption components that allow for an 
acceptable fit.  HD~29419 indicates that additional components are not likely.
  Along this line of sight, the velocity difference between the LIC and Hyades
 Cloud is greatest, and we observe the  maximum Hyades Cloud Mg~II column 
density.  Together, these two characteristics provide two well-defined and 
separated absorption features.  Of course, the other Hyades Cloud lines of 
sight are not as obvious, but are satisfactorily fit by only two components.  
The rest wavelengths and oscillator strengths of the Mg~II lines used in our 
fits are taken from \citet{mort91}, and we use Voigt functions to represent 
the opacity profiles of the fitted interstellar absorption.  The dotted lines 
in Figures 1a, 1b, and 1c are the absorption line fits before convolution with
 the instrumental line spread function.  The thick solid lines in the figures 
that fit the data are the final combined absorption features with the 
instrumental broadening applied.  The instrumental line spread functions 
assumed in our fits are taken from \citet{sahu99}.  

We fit both the h and k lines simultaneously in order to glean all available 
absorption information from the data.  The factor of 2 difference in 
oscillator strengths between the h and k lines can be very useful in 
constraining the interstellar parameters, the stellar continuum level, and the
 number of absorption components.  For example, the Mg~II~k line for HD~27808 
is very saturated, making it difficult to determine unique values for the 
interstellar column density and Doppler parameter.  However, the less opaque 
Mg~II~h line shows a weaker, unsaturated interstellar absorption line.  
Simultaneously fitting both lines allows us to determine the interstellar 
absorption parameters more accurately than by fitting only one of the lines.  
We fit each line separately as well, to see how much the parameters change, 
allowing us to estimate the systematic errors involved in our fits and better 
determine the uncertainties in the various fit parameters.  

Table~3 lists the values for the interstellar absorption parameters and $1 
\sigma$ errors.  For each absorption component there are three parameters: the
 central velocity ($v$ [km~s$^{-1}$]), the Doppler width ($b$ [km~s$^{-1}$]), 
and the Mg~II column density ($\log N_{\rm MgII}$ [$\log$(cm$^{-2}$)]).  The 
central velocity corresponds to the mean projected velocity of the absorbing 
material along the line-of-sight to the star.  If we assume that the warm 
partially ionized material in the solar neighborhood exists in small, 
homogeneous cloudlets each moving with a single bulk velocity, then each 
absorption component will correspond to a single cloudlet, and its velocity is
 the projection of its three-dimensional velocity vector.  The Doppler 
parameter is related to the temperature ($T$ [K]) and nonthermal velocity 
($\xi$ [km~s$^{-1}$]) of the interstellar material by the following equation:
\begin{equation}
b^2 = 0.016629\frac{T}{A}+\xi^2 , 
\end{equation}
where A is the atomic weight of the element in question ($A=24.312$ for Mg).  
Because of the large atomic weight of Mg, the Doppler parameter is more 
sensitive to changes in turbulent velocity than temperature.  The Mg~II column
 density is a measure of the amount of material along the line of sight to the
 star.  If we assume homogeneous cloudlets with constant density, the column 
density will be directly proportional to the cloud thickness.  We have 
identified three separate velocity components, and therefore three individual 
cloudlets, along the line of sight to the Hyades stars.  Columns~2-4 of 
Table~3 list the interstellar parameters for velocity components associated 
with the LIC, columns~5-7 list interstellar parameters for the newly 
identified Hyades Cloud, and columns~8-10 list the same for the third velocity
 component.  The velocity spatial structure is discussed in Section~3.3, the 
Doppler parameter spatial structure is discussed in Section~3.4, and the 
column density spatial structure is discussed in Section~3.5.

\subsection{Spatial Analysis}

The spectra shown in Figures~1a, 1b, and 1c clearly show several distinct 
absorption features.  Because the lines of sight to all of our targets are in 
close proximity, we expect that many of these absorption features should 
sample the same interstellar cloudlet structure.  If the lines of sight to our
 targets pass through the same cloud structures, then the inferred 
interstellar parameters ($b, v, N_{\rm MgII}$) should be similar, and the 
projected spatial structure is probably continuous.  The presence of distinct 
boundaries where the absorption by a cloud is present in one star but not in 
close neighbors would indicate that the cloud has a sharp boundary.  
Furthermore, similar interstellar parameters for the cloud inferred from 
adjacent lines of sight would imply that a cloud is, to first-order, 
homogeneous.

In Table~3 we have grouped the absorption features into three separate 
cloudlets.  The first, present in the spectra of all 18 target stars, has 
velocities ranging from 20-24~km~s$^{-1}$ and is associated with the Local 
Interstellar Cloud (LIC).  The second cloudlet, clearly present in the spectra
 of 10 target stars, has velocities ranging from 12-17~km~s$^{-1}$.  We now 
call this structure the Hyades Cloud (HC).  The third cloudlet, only detected 
toward one target star (HD~28736), has a velocity $\sim$~--4~km~s$^{-1}$.  In 
Figure~2 we plot the spatial distribution of the Hyades target stars, and the 
rough outlines of the Hyades Cloud and the third cloudlet.  Because of the 
limited spatial sampling, the true extent of these clouds is not yet known.  
However, the dashed lines represent the positions between adjacent target 
stars, where the location of the boundary is reasonably well defined.  The 
consistency of the interstellar parameters determined from the spectra of the 
10 target stars, and the contiguous spatial projection, implies that the 
Hyades Cloud is in fact a coherent body with a roughly homogeneous structure.

\begin{figure}
\figurenum{2}
\plotfiddle{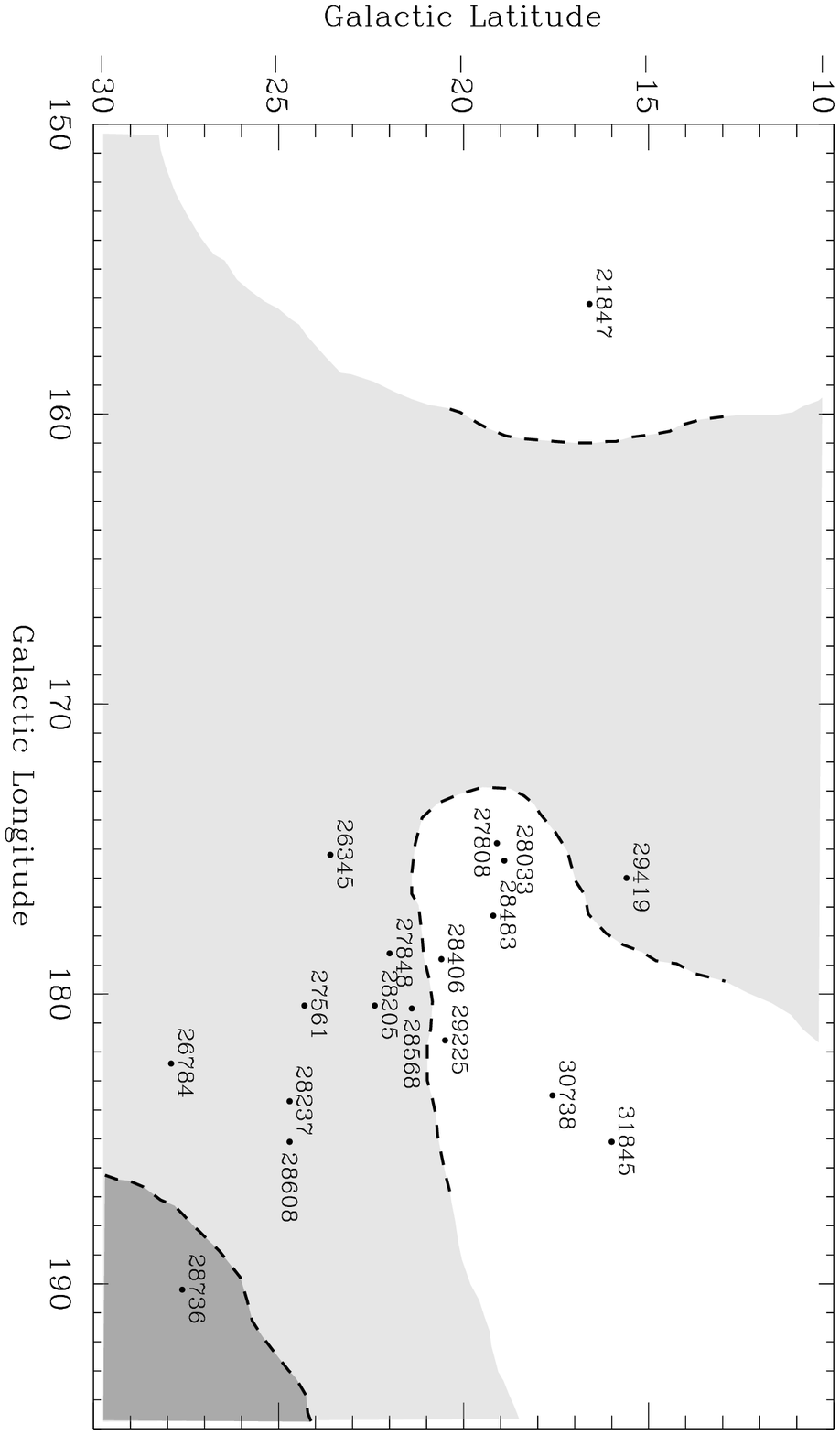}{3.2in}{90}{80}{80}{310}{-50}
\figcaption[fig2.eps]{Locations of the HST Hyades target stars in Galactic 
coordinates.  The dot symbols show the precise location of each Hyades target 
star.  Each symbol is labeled with the HD number associated with the target.  
Information and results for the targets are listed in Tables~2 and 3.  The 
shaded regions represent the tentative locations of the Hyades (light shading)
 and HD~28736 (dark shading) clouds, and dashed lines indicate better 
determined boundaries of the two clouds.  The extent of each cloud in 
directions where no observations exist is obviously tentative. \label{fig2}}
\end{figure}

\placefigure{fig2}

The location of these clouds along the line of sight is a more difficult 
question to answer.  Because absorption due to the LIC has been observed for 
many other stars, \citet{red00} were able to calculate a three-dimensional 
model of the LIC.  Based on this model, the LIC is expected to extend 
$\sim$~5.8~pc in the direction of the Hyades.  The Hyades Cloud could, 
therefore, be located anywhere from the edge of the LIC to the Hyades Cluster,
 perhaps surrounding the target stars themselves.  Figure~3 shows the measured
 Mg~II column density as a function of distance.  If the Hyades Cloud actually
 surrounds the Hyades stars, we would expect to see a trend of increasing 
Mg~II column density with distance.  We see no such trend, and therefore 
conclude that the LIC and Hyades clouds are both located, in their entirety, 
in front of the target stars.  The location of the Hyades Cloud ($d_{HC}$) is 
therefore limited to $5.8~\rm{pc} \leq d_{HC} \leq 41.2~\rm{pc}$.  The mean 
column density for each cloud is given by the dashed line.  For the LIC the 
mean Mg~II column density is $\log N_{\rm MgII}\sim\ 12.63~\log$(cm$^{-2}$).  
The Hyades Cloud has a mean Mg~II column density of $\log N_{\rm MgII}\sim\ 
12.13~\log$(cm$^{-2}$).  The factor of 3 lower mean column density implies 
that the Hyades Cloud is either thinner or less dense than the LIC.  If we 
assume that the Hyades Cloud has similar physical characteristics to the LIC, 
including a similar H~I density ($n_{\rm HI} = 0.10~\rm{cm}^{-3}$) and Mg~II 
depletion (--1.1 dex), then we estimate the thickness of the Hyades Cloud to 
be $\sim\ 1.8~\rm pc$.  

\begin{figure}
\figurenum{3}
\plotfiddle{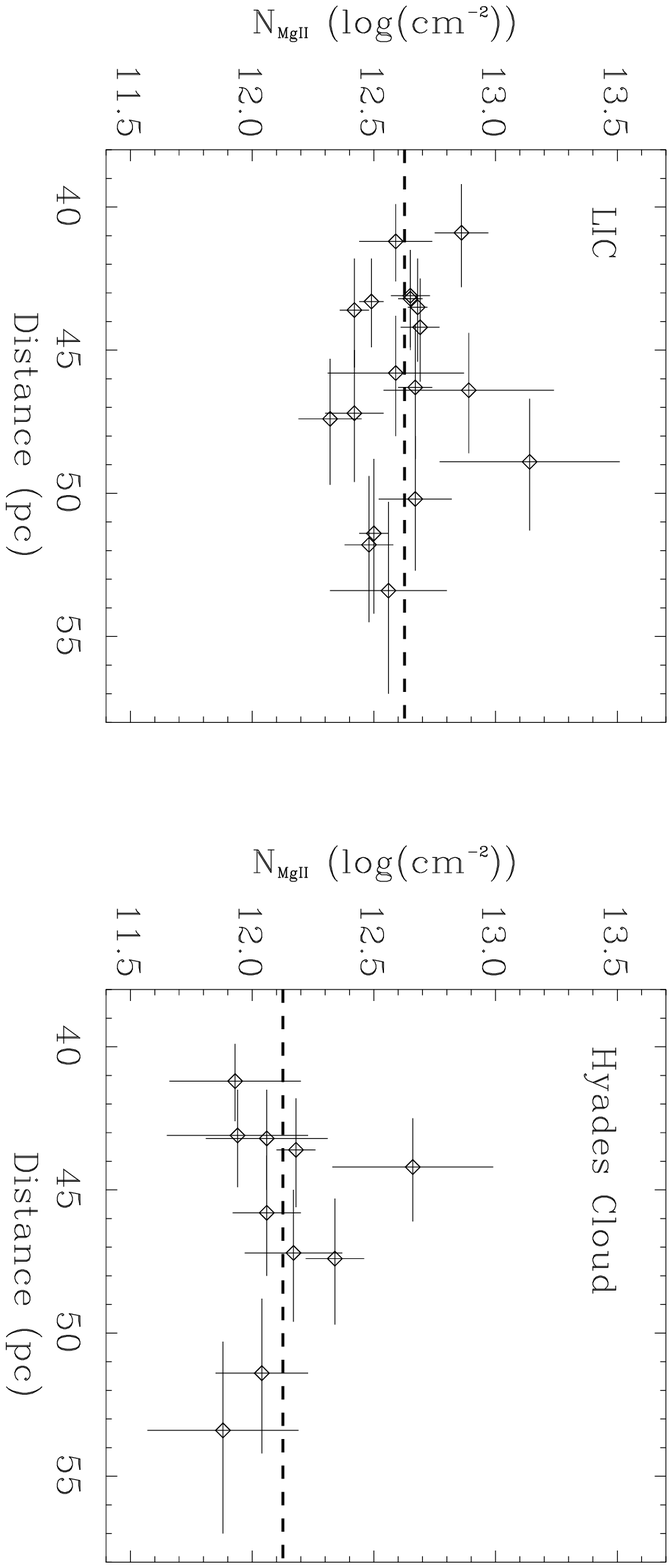}{3.in}{90}{80}{80}{310}{0}
\figcaption[fig3.eps]{Variation of Mg~II column density with heliocentric 
distance, along with 1$\sigma$ errors.  The mean Mg~II column density is 
portrayed by the thick dashed line.  The left panel displays the Mg~II column 
densities associated with the LIC.  The right panel displays the Mg~II column 
densities associated with the Hyades Cloud.  Because there is no tendency in 
the Mg~II column densities to increase with increasing distance, we conclude 
that the LIC and Hyades clouds are located in front of the target stars and do
 not surround them.\label{fig3}}
\end{figure}

\placefigure{fig3}

\subsection{Velocity Structure}

Since we can fit the LISM absorption successfully with Voigt profiles, we 
presume that the medium along a given line of sight for each cloud is moving 
at a single bulk velocity,  with symmetric thermal and turbulent motions.  The
 measured central velocity and 1$\sigma$ error bars for each absorption 
feature is given in Table~3.  The velocities of absorption features that we 
identify with the LIC are given in column~2, those that we identify with the 
Hyades Cloud are given in column~5, and the third component is given in 
column~8.  

If we assume that each cloudlet structure in the LISM is moving with a 
single-valued bulk velocity, then we can calculate a unique velocity vector 
for each individual cloud.  Due to the small number of lines of sight 
previously sampled, unique velocity vectors have only been calculated for two 
clouds: the LIC and the Galactic (G) Cloud \citep{lall95}.  In the 
heliocentric rest frame, the LIC, in which the Sun is embedded, is flowing 
toward Galactic coordinates $l = 186^{\circ}$ and $b = -16^{\circ}$ at a speed
 of $25.7~\pm~0.5$~km~s$^{-1}$ \citep{witte93, lall92}.  Using this vector we 
compute the projected LIC velocities ($v_{\rm{LIC}}$), for the lines of sight 
toward our Hyades stars.  In Table~4 the predicted velocity values are given 
in column~2, and the observed central velocities associated with the LIC are 
given in column~3.  We observe LIC absorption in all 18 Hyades target stars, 
but the observed velocity values are systematically lower than the predicted 
values.  The mean difference ($v_{\rm predicted}(\rm LIC) - 
v_{\rm observed}(\rm LIC)$) is $2.9~\pm~0.7$~km~s$^{-1}$.  Two plausible 
explanations for this discrepency are that unresolved blended absorption 
features are offsetting our central velocity fits, or that the calculated LIC 
velocity vector is inaccurate for this region of the sky.  The presence of the
 Hyades Cloud absorption (typically $\sim\ 7.8$~km~s$^{-1}$ from the observed 
LIC velocity) does not seem to affect the LIC velocity measurements.  The 10 
target stars that show Hyades absorption, often blended with the LIC 
absorption, have a mean LIC velocity difference similar to the overall mean 
($v_{\rm predicted}(\rm LIC) - v_{\rm observed}(\rm LIC) = 2.7 \pm 
0.8$~km~s$^{-1}$), as do those without Hyades absorption features ($v_{\rm 
predicted}(\rm LIC) - v_{\rm observed}(\rm LIC) = 3.1 \pm 0.6$~km~s$^{-1}$).  
The LIC velocity vector determined by \citet{lall92} used nine lines of sight 
distributed over the entire sky.  The nearest target to the Hyades stars in 
their sample is $\eta$~Aur, about 25$^{\circ}$ away, with absorption at 
$23.0~\pm~1.5$ km~s$^{-1}$.  Now that many more lines of sight have been 
studied since the \citet{lall92} determination of the LIC velocity vector, a 
revision of the vector could bring the Hyades velocities into better 
agreement.  However, the $2.9~\pm~0.7$~km~s$^{-1}$ discrepancy between the 
predicted and the observed projected LIC velocity cannot be accounted for by 
small corrections in the LIC velocity vector direction or magnitude.  
Therefore, a single bulk velocity flow may not be realistic throughout the 
whole LIC.  Note that the leading edge of the LIC, moving with a velocity of 
25.7 km~s$^{-1}$ along the line of sight towards $l = 186^{\circ}$ and $b = 
-16^{\circ}$, is very near to the lines of sight to the Hyades stars.  Our 
observations of slightly slower velocities may indicate a compression and 
deceleration at the leading edge of the LIC.  We will address these questions 
in a future paper.

\begin{figure}
\figurenum{4}
\plotfiddle{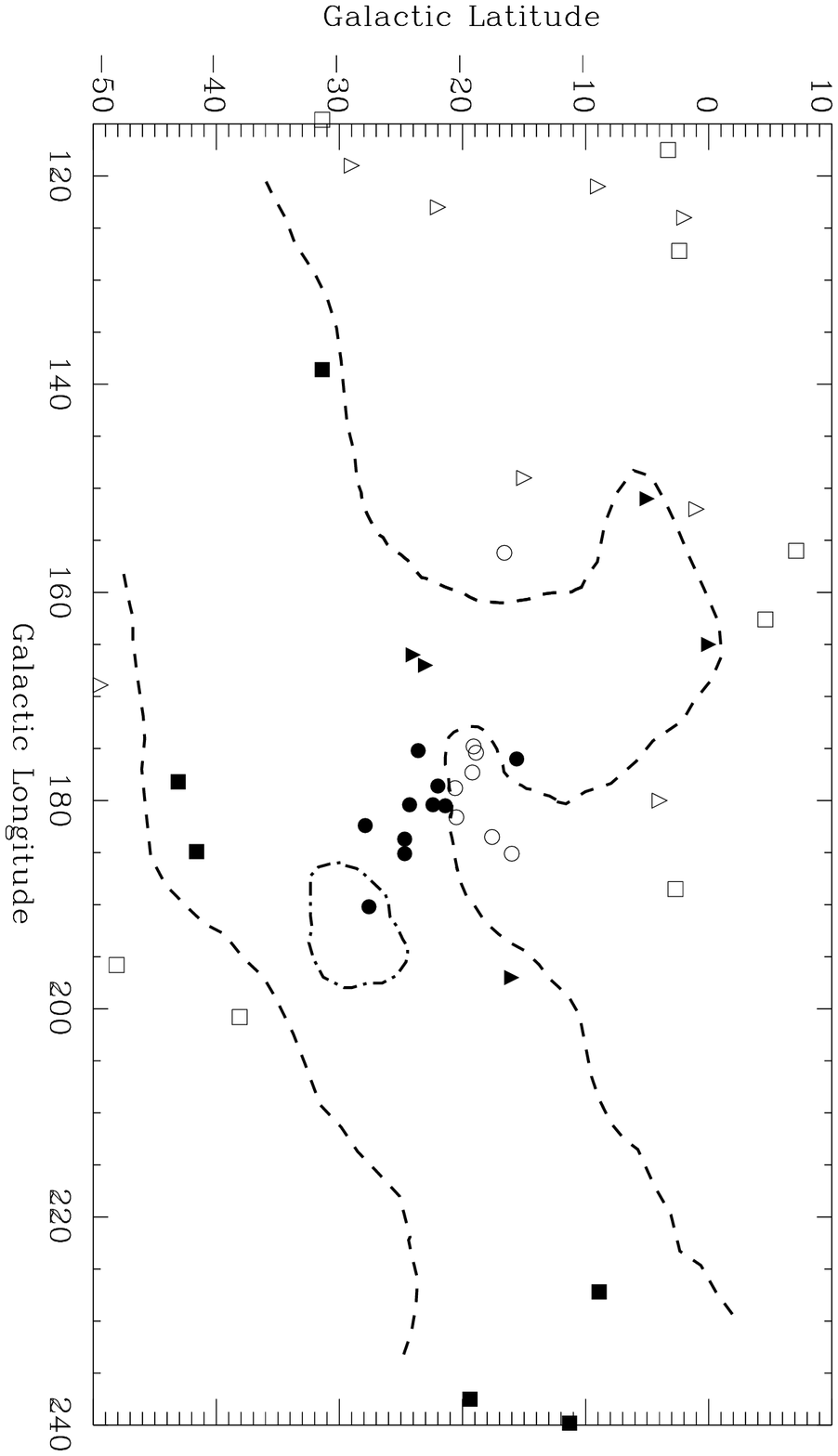}{3.2in}{90}{80}{80}{310}{-50}
\figcaption[fig4.eps]{Locations, in Galactic coordinates, of HST Hyades target
 stars and other stars observed in the UV with HST and in the optical with 
ground-based instruments.  Lines of sight that show absorption at velocities 
consistent with the Hyades Cloud (10-20 km~s$^{-1}$) are shown as filled 
symbols and those that do not are shown as open symbols.  The Hyades stars are
 identified by circles.  HST observations of other stars are displayed as 
squares.  Other stars with ground-based Ca~II observations are displayed as 
triangles.  The dashed line is a tentative outline of boundary of the Hyades 
Cloud.  The dashed-dotted line is a tentative outline of the boundary of the 
third cloudlet observed toward HD~28736.  No other targets show absorption at 
the velocity of this cloud.  \label{fig4}}
\end{figure}

\placefigure{fig4}

Ten of the 18 Hyades Stars show a second absorption feature in the velocity 
range of 12-17~km~s$^{-1}$ with a mean value of 14.8~km~s$^{-1}$.  We have 
identified what we now call the Hyades Cloud based on the similar absorption 
velocities measured for these ten lines of sight.  Because the Hyades stars 
cover a small area of the sky, and do not show a clear trend in velocity 
across the sample area, it is not yet feasible to calculate a velocity vector.
  If the kinematic structure of the LISM is coherently organized, then the 
direction of the Hyades Cloud velocity vector is expected to be similar to the
 direction of the LIC velocity vector.  The G Cloud and the LIC vectors, for 
example, are very similar \citep{lall95}.  Therefore, we anticipate that other
 nearby targets with lines of sight that also traverse the Hyades Cloud, will 
have absorption at velocities $<$~5~km~s$^{-1}$ from the measured mean value 
of Hyades absorption at 14.8~km~s$^{-1}$.  Figure~4 shows an expanded view of 
the Hyades region in Galactic coordinates with other nearby stars plotted 
along with the Hyades stars.  The Hyades sample is plotted as circles, while 
other targets that have Mg~II absorption detected by HST are plotted as 
squares, and targets that have Ca~II absorption observed with ground-based 
telescopes are plotted as triangles.  Only high resolution data 
(R $= \lambda/\Delta\lambda\gtrsim 100,000$) are included in these samples.  
The HST targets are listed in Table~5 and the Ca~II targets are listed in 
Table~6.  We have separated those stars that show absorption in the velocity 
range from 10-20 km~s$^{-1}$ from those that do not.  In Figure~4, those 
targets that show absorption at a velocity consistent with the Hyades Cloud 
are plotted as filled symbols, while those that do not are plotted as open 
symbols.  This provides a tentative means for establishing the boundary of the
 Hyades Cloud.  Clearly the Hyades Cloud does not extend far to the Galactic 
North-East (upper left in the figure), due to the large number of objects that
 do not show Mg~II or Ca~II absorption at the Hyades Cloud velocity.  Nor does
 the Hyades Cloud extend far to the Galactic South-West (lower right in the 
figure), as $\epsilon$~Eri and 40~Eri~A do not show any Hyades Cloud 
absorption.  It appears that the Hyades Cloud structure may extend almost 
linearly across the sky from Galactic South-East to Galactic North-West, in a 
filamentary structure.  Definitive proof of Hyades Cloud absorption can only 
be made by self-consistent measurements of a velocity vector, cloud 
temperature, and elemental abundances.  As this requires high resolution 
spectra in the 1200-1350~\AA\ region, which are not yet available, this task 
is beyond the scope of this paper, and we simply provide a tentative boundary 
based on similarity of absorption velocity.  The large apparent size of the 
Hyades Cloud suggests that it is probably located nearby, just beyond the LIC.
  The fact that Sirius, only 2.6~pc away, may be a member of the Hyades Cloud,
 supports the indication that the Hyades Cloud is nearby.  The absorption at 
the Hyades Cloud velocity towards Sirius was identified by \citet{lall94} as 
the Blue Cloud.

In Figure~4, a tentative boundary (the dotted-dashed line) is given for the 
third component that is observed towards HD~28736.  No other nearby stars show
 absorption at a velocity similar to the third component velocity 
(--4.3 km~s$^{-1}$).  Therefore, this third component cloud is limited to a 
small region on the sky, indicating that it is likely located farther away 
than the Hyades Cloud.

\subsection{Temperature and Turbulent Structure}

The observed value of the Doppler parameter ($b$) is a measure of the 
temperature ($T$) and turbulent velocity ($\xi$).  The relationship is 
provided in Equation~1.  Because of its relatively large atomic weight, 
magnesium's Doppler parameter is primarily due to turbulence or unresolved 
clouds along the line of sight.  The measured Doppler parameter and 1$\sigma$ 
error bars for each absorption feature are given in Table~3.  If we assume 
that the temperature and turbulent structure in a cloud are relatively 
constant, the Doppler parameters should be equal.  For the LIC absorption 
component, all 18 lines of sight are in good agreement regarding the Doppler 
parameter.  The mean b-value for the LIC is 
$b_{\rm LIC}~3.5~\pm$~0.6~km~s$^{-1}$.  If we assume a LIC temperature of 
$T_{\rm LIC}~\sim~8000~\pm~1000~\rm{K}$ \citep{pisk97,lin95,lall94,witte93}, 
then the turbulent velocity in this region of the LIC is 
$\xi_{\rm LIC}~\sim~2.6~\pm~0.8$~km~s$^{-1}$.  This value is larger than 
values calculated for the LIC along other lines of sight 
\citep{pisk97,lall94,lin95}, but not for the LISM in general 
\citep{lall94,wood00}.  Since the mean $b_{\rm LIC}$ is identical for those 
lines of sight with and without additional Hyades absorption, we do not 
believe that unresolved blends are compromising our measurements of the 
turbulence values.

The weaker, blended Hyades absorption component permits us to determine the 
Doppler parameter but with lower precision.  The 1$\sigma$ error bars are 
larger owing to the absorption being a blended and weaker line.  The mean 
Mg~II b-value for the Hyades component is 
$b_{\rm HC}=2.8~\pm$~0.9~km~s$^{-1}$, indicating a lower temperature cloud or 
a less turbulent structure than the LIC.  In Table~5, the other Mg~II 
observations of nearby targets give b-values that agree within the errors of 
the mean Hyades sample b-value.  This is an additional indication that these 
lines of sight traverse the same cloud structure.

Many of the stars observed by HST and listed in Table~5 that show Hyades Cloud
 absorption in Mg~II also show Hyades Cloud absorption in other atomic lines. 
 Observations of absorption in many lines allow the temperature and turbulent 
velocity in the cloud to be measured.  These parameters are listed in 
columns~9 and 10 of Table~5.  The measured b-values of Mg~II and H~I Hyades 
Cloud absorption along the line of sight to HR~1099 do not permit a unique 
determination of the temperature or turbulent velocity.  Observations of 
Hyades Cloud absorption in other atomic lines are needed to satisfactorily 
determine these two quantities along this line of sight.  Likewise, the 
complicated LISM absorption spectra seen in observations of $\beta$~CMa led 
\citet{dup98} to force a Hyades Cloud absorption component fixed by the ISM 
parameters measured towards $\epsilon$~CMa.  Although the absorption toward 
$\beta$~CMa is therefore consistent with Hyades Cloud absorption, the 
$\beta$~CMA data provide no additional information on the physical properties 
of the Hyades Cloud.  Therefore, the interstellar parameters for $\beta$~CMa 
are not listed in Table~5, and those interstellar parameters listed for 
HR~1099 should be given low weight.  

\subsection{Column Density Structure}

The observed Mg~II column density indicates the number of Mg$^{+}$ ions along 
the line of sight.  If we assume a homogeneous medium with a constant density,
 the column density is directly proportional to the distance that the line of 
sight traverses through the cloud.  The measured Mg~II column densities and 
1$\sigma$ error bars for each absorption feature are given in Table~3.  
Column~4 lists the Mg~II column density associated with the LIC.  Figure~5 
shows a crude contour plot of the LIC Mg~II column density.  A negative 
gradient in column density is clearly evident from Galactic North-East to 
Galactic South-West.  The target with the maximum column density is HD~21847 
($l = 156.2$ and $b = -16.6$) and the minimum column density through the LIC 
is in the line of sight of HD~26784 ($l = 182.4$ and $b = -27.9$).  The 
gradient is fairly smooth between these two stars.  \citet{red00} developed a 
three-dimensional model of the macroscopic structure of the LIC using 32 lines
 of sight, and assuming a homogeneous, constant density cloud structure.  
Figure~5 shows the H~I column density predicted from their model as shading.  
The model shows a gradient in H~I column density that is similar to what is 
seen in Mg~II over the same region.  The predicted values of the H~I column 
density for the lines of sight to the Hyades stars are listed in column~4 of 
Table~4.  Although the depletion of Mg may change somewhat for different lines
 of sight through the LIC (see the next section), we do not expect the Mg 
depletion to change drastically over small distances.  Our result that the two
 column densities are indeed found to be consistent is an excellent check for 
the \citet{red00} LIC model, and for the quality of the Hyades analysis. 

\begin{figure}
\figurenum{5}
\plotfiddle{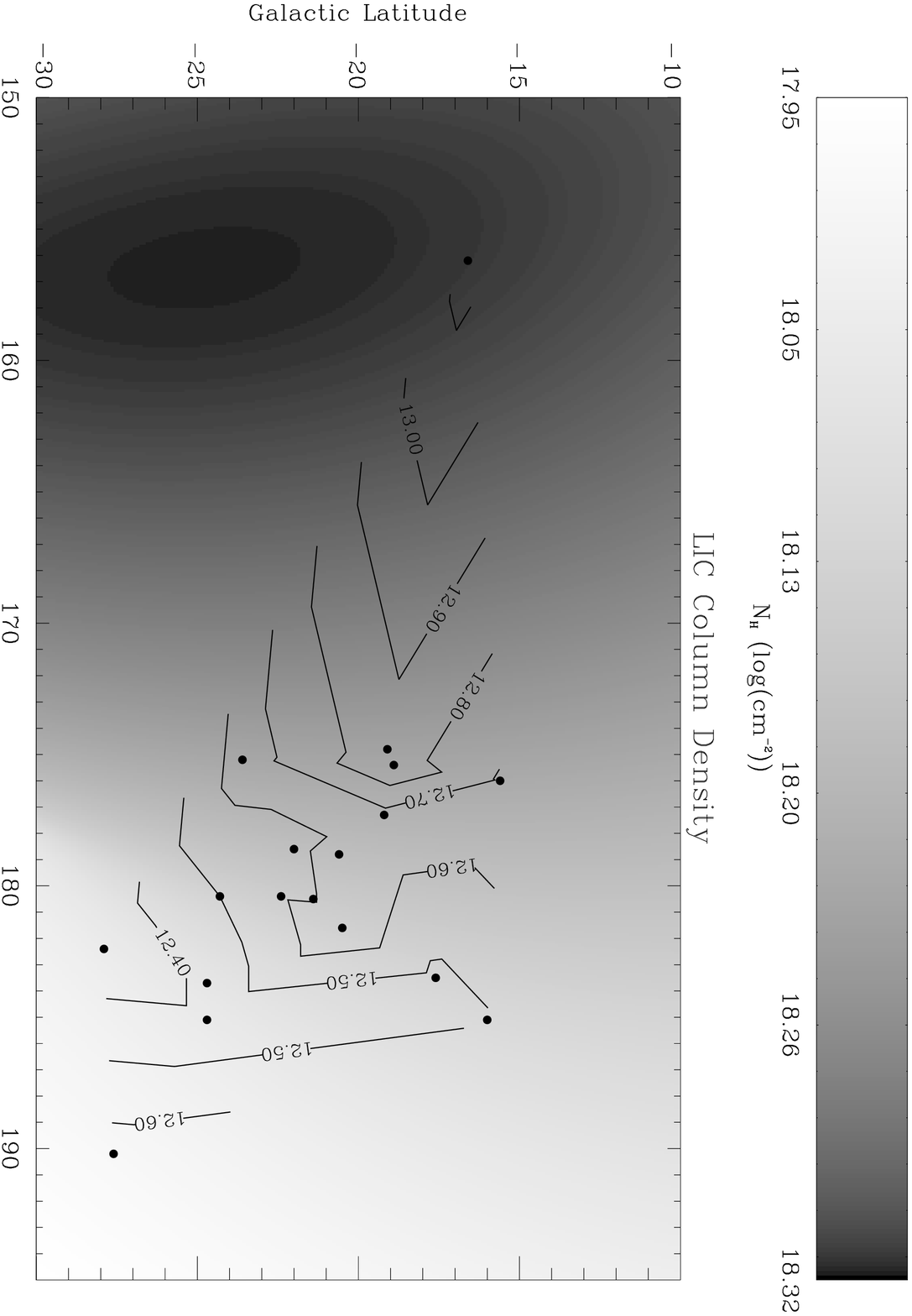}{5.in}{90}{75}{75}{310}{20}
\figcaption[fig5.eps]{Contours of the measured Mg~II column density and 
shading for the predicted H~I column density for the LIC.  The filled circle 
symbols ($\bullet$) display the location of the Hyades stars in Galactic 
coordinates, identical to Figure~2.  The labelled lines represent contours of 
measured Mg~II column density.  The shaded gradient represents the H~I column 
density predicted by the LIC model of \citet{red00}.  The range of H~I column 
densities is given by the bar at the top of the figure.  Note that the 
gradient of Mg~II column density is similar to the H~I column density 
gradient.  \label{fig5}}
\end{figure}

\begin{figure}
\figurenum{6}
\plotfiddle{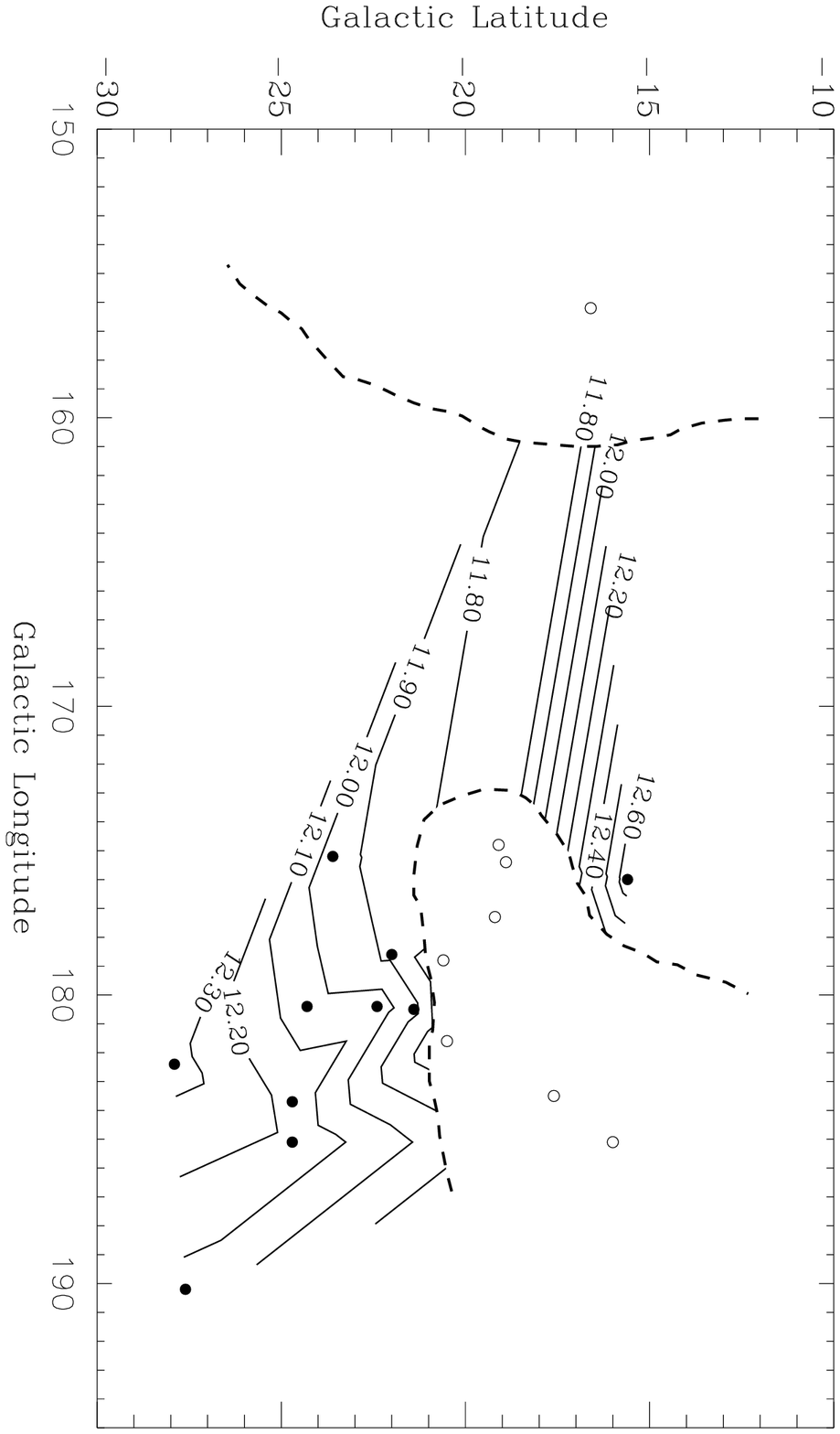}{3.2in}{90}{80}{80}{310}{-50}
\figcaption[fig6.eps]{Contours of the measured Mg~II column density for the 
Hyades Cloud.  The filled circle symbols ($\bullet$) display the location of 
the Hyades stars in Galactic coordinates, identical to Figure~2.  The labelled
 lines represent contours of measured Mg~II column density.  Our tentative 
outline of the Hyades Cloud is given by the dashed lines.  Among the main 
group of Hyades Cloud lines of sight (those lines of sight in the South-west 
portion of the figure), there is a clear gradient of lower Mg~II column 
densities toward the cloud's edge.  There may still be Hyades Cloud absorption
 beyond the boundary, but at levels less than our detection limit.  
\label{fig6}}
\end{figure}

\placefigure{fig5}
\placefigure{fig6}

As discussed in Section~3.2, the mean Mg~II column density through the Hyades 
Cloud is more than a factor of 3 smaller than the Mg~II column density through
 the LIC.  The measured values of the Hyades Cloud Mg~II column density are 
listed in column~7 of Table~3.  Due to the random error associated with the 
low S/N of the observations, and the systematic errors created by the 
uncertainty in the continuum and the tendency for the Hyades Cloud absorption 
to be blended with the LIC absorption, we estimate that the minimum Mg~II 
column density that could be detected to be 
$\log~N_{\rm MgII}=11.5~\log(\rm{cm}^{-2})$.  This value includes the random 
errors of measurement and our estimate of the more important systematic 
errors.  The upper limit of the Mg~II column density for our Hyades Cloud 
nondetections is therefore $\log~N_{\rm MgII}=11.5~\log(\rm{cm}^{-2})$.  Like 
the LIC, the Hyades Cloud shows a smooth gradient in column density with 
position on the sky.  Figure~6 shows the Mg~II column density contours for 
those lines of sight that show absorption greater than our detection limit, 
consistent with Hyades Cloud velocities.  The tentative outline of the Hyades 
Cloud from Figure~2 is also provided.  The maximum Mg~II column density is 
exhibited by the isolated star HD~29419, and the minimum detected Mg~II 
column density is for the line of sight to HD~27848, located close to the edge
 of the Hyades Cloud at $l = 178.6$ and $b =  -22.0$.  The Mg~II column 
density gradient clearly decreases from the interior of the Hyades Cloud 
towards the edge near HD~27848.  Other nearby stars that show absorption at 
velocities similar to the Hyades show Mg~II column densities consistent with 
those found in the Hyades sample itself.  These values are given in Table~5, 
and provide additional evidence that these lines of sight traverse the same 
physical cloud as the Hyades sample.  

\subsubsection{Mg~II Depletion}

The gas-phase abundance of elements in the LISM is an important physical 
parameter for our understanding of the structure of the LISM.  The analysis of
 \citet{lall94} demonstrated that Mg~II is the dominant ionization state of 
magnesium in the LISM.  It has been found that the gas-phase abundance of 
Mg~II can vary considerably in the LISM \citep{pisk97, dring97}.  Depletion is
 defined by the equation,
\begin{equation}
D(\rm{Mg}) = \log\left[\frac{N_{\rm MgII}}{N_{\rm HI}}\right] - 
\log\left[\frac{\rm{Mg}}{\rm{H}}\right]_{\odot},
\end{equation}
where the solar abundance of magnesium to hydrogen is 
$\log(\rm{Mg/H})_{\odot} = -4.41$ \citep{and89}.  Note that these depletion 
values assume that all hydrogen in the LISM is neutral.  The ionization state 
of the LISM is uncertain, but it is likely that nearly half of the hydrogen is
 ionized \citep{wood97,lall94}.  In this case, the depletion values calculated
 from Equation~2 may be too large by 0.3 dex or more.  Depletion values 
derived by \citet{pisk97} and \citet{dring97} imply a LIC depletion value for 
magnesium of D(Mg)~$=~-1.1~\pm~0.2$.  This is similar to the depletion 
D(Mg)~$=~-0.89~\pm~0.05$ obtained for the warm cloud in the line of sight to 
$\zeta$~Oph \citep{sav96}.

The Hyades sample unfortunately does not have corresponding spectra of H~I 
(1216~\AA), and therefore does not allow for an independent measure of the 
hydrogen column density and magnesium depletion.  However, the \citet{red00} 
LIC model provides predictions of the H~I column densities through the LIC for
 any line of sight.  In Table~4 the predicted LIC hydrogen column densities 
are listed in column~4, and the measured LIC Mg~II column densities, as taken 
from Table~3, are listed in column~5.  The calculated depletions are listed in
 column~6.  The Mg depletions range from $D(\rm{Mg}) = -0.76~\pm~0.37$ to 
$D(\rm{Mg}) = -1.32~\pm~0.13$.  Therefore all of the Hyades stars are 
consistent with the canonical value of LIC depletion, $D(\rm{Mg}) = 
-1.1~\pm~0.2$.  This provides further confirmation that not only do the 
predicted H~I column densities have a gradient similar to that of the measured
 Mg~II column densities, but the difference between the two provides a 
reasonable estimate of the magnesium depletion.  

The magnesium depletions in the Hyades Cloud measured by HST observations of 
nearby stars are listed in column~11 of Table~5.  There is no clear trend 
among the magnesium depletion values, but they are roughly similar to the LIC 
value of $D(\rm{Mg}) = -1.1~\pm~0.2$.  We note that the two stars with values 
of D(Mg) furthest from the LIC value, HR~1099 and $\epsilon$~CMa, have three 
or more velocity components in their lines of sight and therefore, the 
hydrogen column density and magnesium depletion are difficult to measure for 
the Hyades Cloud component.

\section{Homogeneity in the LISM}

The close proximity of the lines of sight to the Hyades stars provides an 
excellent opportunity to investigate the small-scale microstructure of warm 
interstellar clouds.  Due to the large number of targets observed in the 
Hyades sample (18), and the small angular distance between them, there are 
large numbers of unique pairs with which to compare observed ISM parameters.  
Because the LIC is observed in all 18 targets, there are 153 unique pairings 
of targets varying in angular distance from $0.6^{\circ}$ to $33.2^{\circ}$.  
Figure~7 shows the distribution of angular distances of the Hyades stars in 
$1^{\circ}$ bins.  The left panel gives the distribution for the LIC, and it 
is clear that the vast majority of star pairings are less than $10^{\circ}$ 
away from each other.  The same is true for the Hyades Cloud, as shown in the 
right panel of Figure~7.  However, there are only 45 unique pairs for the 10 
stars with Hyades Cloud absorption.  The Hyades Cloud distribution of angular 
separations range from $1.0^{\circ}$ to $17.8^{\circ}$.

\begin{figure}
\figurenum{7}
\plotfiddle{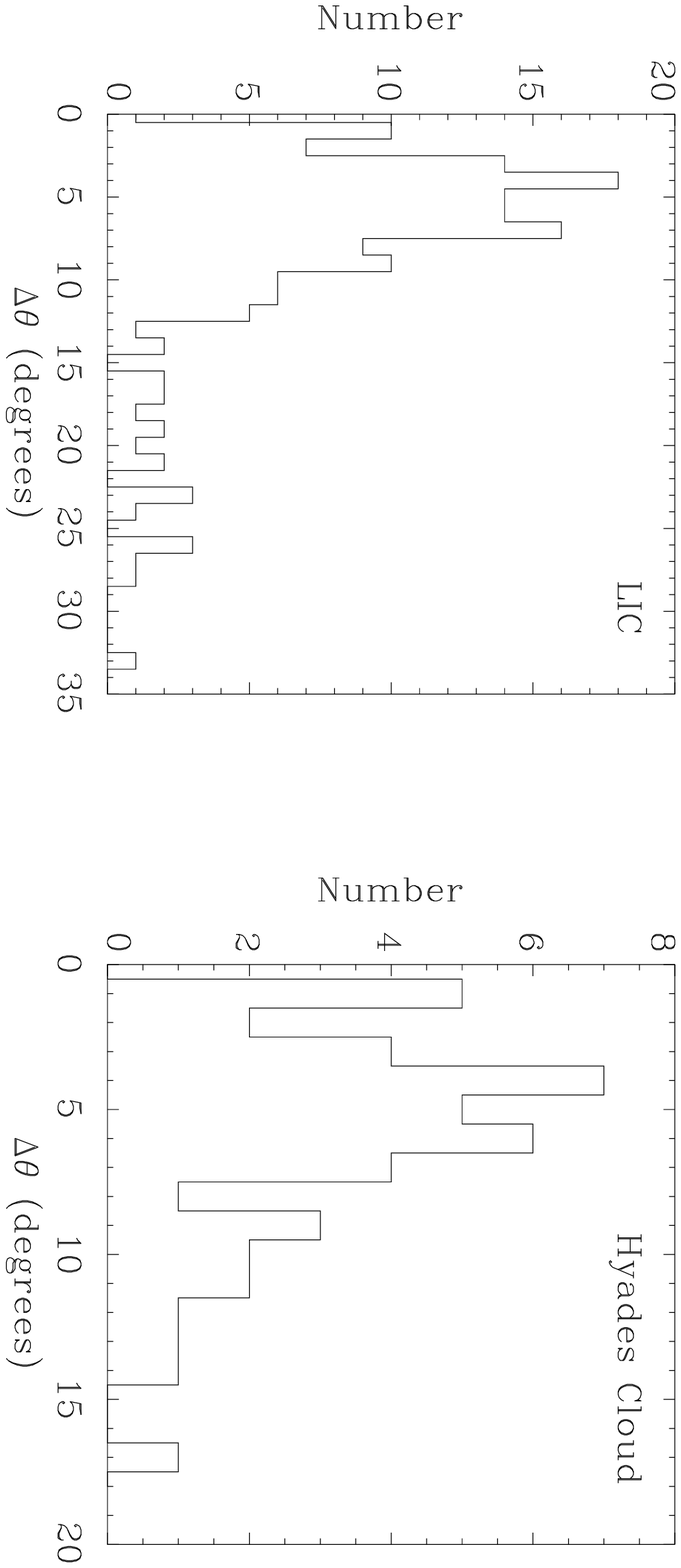}{3.in}{90}{80}{80}{310}{0}
\figcaption[fig7.eps]{Distribution of angular distances between Hyades stars 
in $1^{\circ}$ bins.  The left panel displays the distribution of all the 
Hyades target stars.  There are 18 targets in all, and therefore 153 unique 
pairs.  The right panel displays the distribution of all Hyades target stars 
that show Hyades Cloud absorption.  Ten targets have lines of sight that 
traverse the Hyades Cloud, and there are therefore 45 unique pairs.  Both 
distributions show that the majority of the pairs are separated by less than 
10 degrees.  \label{fig7}}
\end{figure}

\placefigure{fig7}

We compare the observed Mg~II column density for the LIC and the Hyades Cloud 
as a function of angular separation for each individual pairing.  This 
comparison should provide some insight into the density inhomogeneity of the 
LISM.  If there are large inhomogeneities, proximate lines of sight should not
 show similar column densities.  In fact, we find that close lines of sight do
 show similar Mg~II column densities.  Figure~8 shows the difference in the 
observed Mg~II column density for all pairings as a function of angular 
separation.  The closest pairs in fact show excellent agreement in observed 
column densities.  The top panel displays the observed Mg~II column densities 
of the LIC for all 18 target stars.  The bottom panel displays the observed 
Mg~II column densities for the ten stars that show Hyades Cloud absorption.  
The similar values of $N_{\rm MgII}$ for the closest pairs indicates that the 
LIC and Hyades Cloud are in fact roughly homogenous.  For the LIC, we have 
some indication of the distance to the edge of the cloud from a model based on
 previous observations of 32 lines of sight.  \citet{lin00} presented a 
methodology for calculating the distance to the edge of the LIC, using high 
resolution observations of deuterium absorption and assuming constant D/H 
ratio in the LIC.  \citet{red00} used this technique to calculate distances to
 the edge of the LIC and to fit a three-dimensional model of the shape of the 
LIC.  We use these estimates to the edge of the LIC to calculate physical 
separations at the edge of the LIC for different sight lines from the angular 
separations of the Hyades stars.  In the top panel of Figure~8, the top axis 
is this distance scale in parsecs.  The lines of sight of the very closest 
pairs of targets, which show excellent agreement, are on the order of 0.05-0.1
 pc away from each other at the edge of the LIC, corresponding to separations 
of (1-2)~$\times 10^4$~AU.  One would not be surprised to see differences in 
column density to increase with increased distance, due to the macroscopic 
characteristics of the cloud, but it is important to determine whether the 
column densities change smoothly or stocastically with increasing angular 
separations.  We have indicated where a factor of two in $\Delta~N_{\rm MgII}$
 occurs in Figure~8.  The change in column density for pairs of targets does 
not consistently exceed a factor of 2 until angular distances 
$\gtrsim~8^{\circ}$ or physical distances of $\gtrsim$~0.6~pc are obtained.  
The transition to larger $\Delta~N_{\rm MgII}$ for pairs of stars seems to 
occur smoothly.

\begin{figure}
\figurenum{8}
\plotone{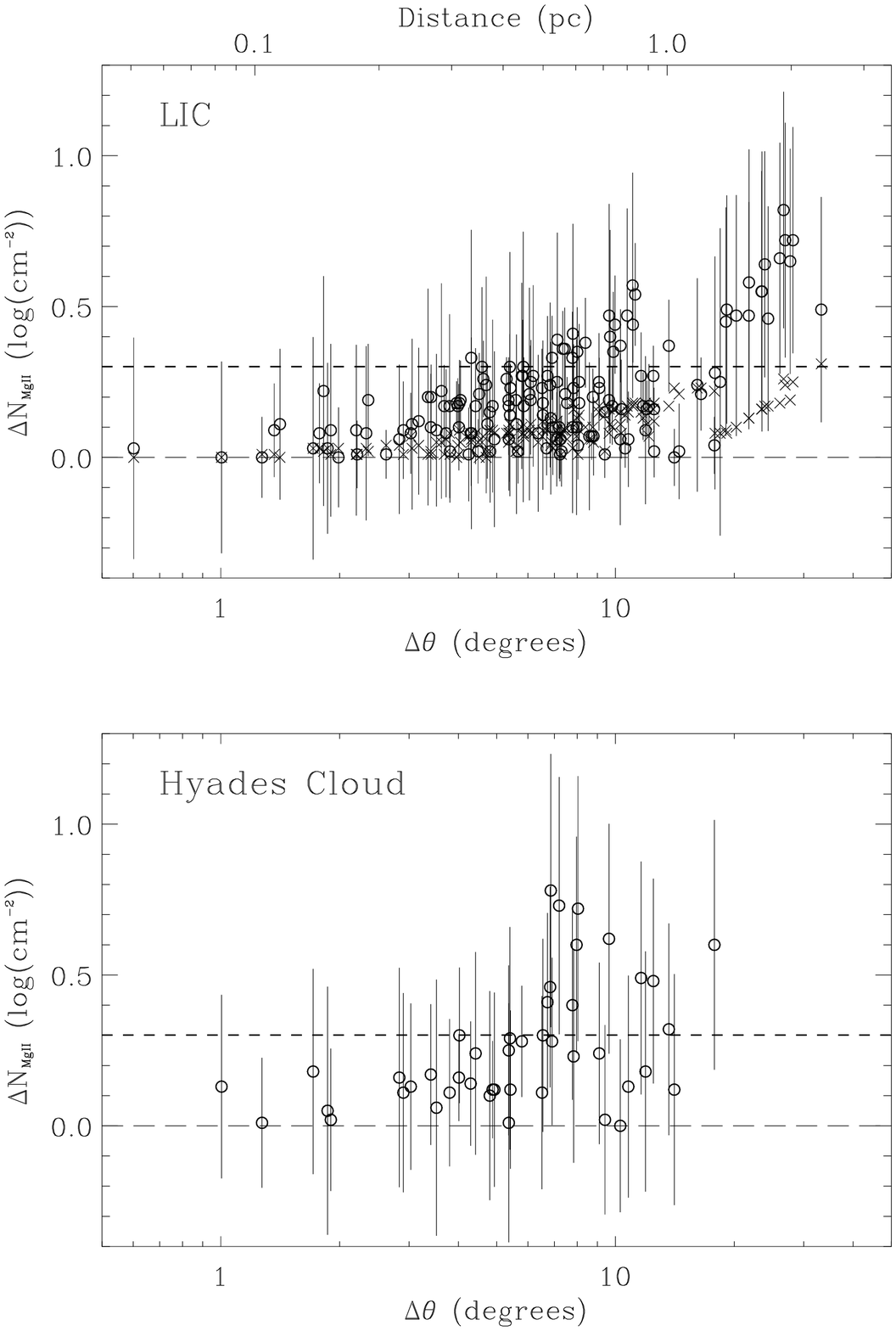}
\figcaption[fig8.eps]{\label{fig8}}
\end{figure}

\placefigure{fig8}

In the top panel of Figure~8 we also include the change in H~I column density,
 $\Delta~N_{\rm HI}$, as predicted by the \citet{red00} macroscopic model of 
the LIC.  These calculations are displayed by the cross-hatch symbols 
($\times$).  Clearly, the changes in the predicted H~I column density between 
lines of sight to the Hyades stars are smaller than those of the measured 
Mg~II column density.  This indicates that the LIC, as measured by the Mg~II 
column densities of this Hyades sample, is more inhomogeneous than the model. 
 The most likely explanation is that the LIC model is undersampled, and 
requires many more lines of sight in order to resolve the column density 
flucuations.  Our sample indicates that at angular separations of 
$\lesssim~8^{\circ}$ the measured column densities in the LIC do not change by
 more than a factor of two.

\begin{figure}
\figurenum{9}
\plotone{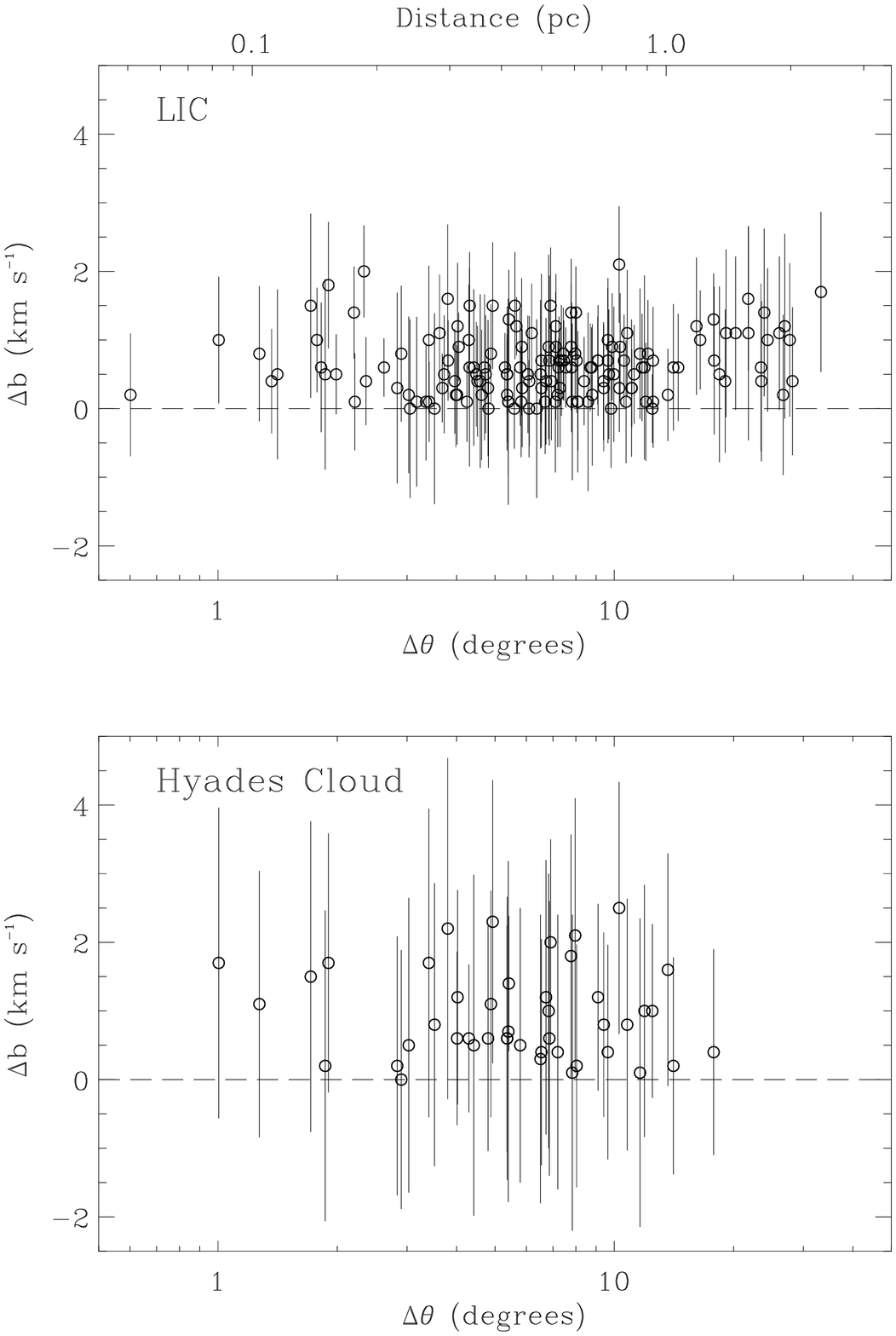}
\figcaption[fig9.eps]{\label{fig9}}
\end{figure}

\placefigure{fig9}

We can also study the inhomogeneity of the Doppler parameter.  In Section~3.4 
we discussed how the Doppler parameter is related to the temperature and 
turbulent velocity structure.  With the many lines of sight to the Hyades 
stars, we can question whether the LIC and the Hyades Cloud have constant 
temperatures and turbulent velocities, or whether these quantities change at 
boundaries or interfaces.  In Figure~9 we display the difference in the 
Doppler parameter for all pairs of stars for both the LIC and the Hyades 
Cloud.  Clearly, the Hyades Cloud shows larger error bars due to the weak and 
blended absorption lines.  The LIC, on the other hand, has relatively small 
error bars.  There appears to be a distinct contrast in the difference of 
Mg~II column density and Doppler parameter data.  Whereas the Mg~II column 
density differences show a gradual increase in difference with increasing 
distance, the Doppler parameter differences do not show any trend with 
distance, but rather a constant difference of $\sim~1$~km~s$^{-1}$ for all 
distances, consistent with the uncertainties in the measurements.   It is 
possible that the spectral resolution of our observations is not high enough 
to adequately determine the Doppler parameter with great enough precision to 
detect variations in $b$.  However, if we are successfully resolving and 
fitting the Doppler parameter for the observed absorption features, this 
behavior of the Doppler parameter may indicate that there are no large 
gradients in temperature or turbulent velocity.  

\section{Conclusions}

We have analyzed Mg~II h and k absorption features in HST/STIS spectra for 18 
members of the Hyades Cluster.  Our findings are summarized as follows:
\begin{enumerate}
\item Three velocity components are observed along the lines of sight toward 
the Hyades Cluster.  All stars show absorption that is consistent with the LIC
 at $\gtrsim~20$~km~s$^{-1}$.  Ten of the 18 stars exhibit a secondary 
component at radial velocities of 12-16~km~s$^{-1}$.  One star (HD~28736) 
shows absorption of a third component at a velocity of $-4.3$~km~s$^{-1}$. 
\item The observed radial velocities associated with the LIC absorption are 
lower than those predicted by the LIC velocity vector \citep{lall92,lall95} by
 $2.9~\pm~0.7$~km~s$^{-1}$.  The direction to the Hyades Cluster is also 
roughly the direction of the leading edge of the LIC, so the lower velocities 
may indicate a compression or deceleration of material at the leading edge of 
the LIC. 
\item The Hyades Cloud appears to extend beyond the field of view of the 
Hyades, as other nearby lines of sight show absorption at similar velocities. 
 We have collected data from previous HST observations and ground-based Ca~II 
observations.  The indicated structure of the Hyades Cloud is more filamentary
 in nature than that inferred for the LIC.  
\item The large number of proximate lines of sight provides a unique 
opportunity to investigate the inhomogeneity of the LISM, and in particular 
individual clouds such as the LIC.  We find that the range in Mg~II column 
densities does not exceed a factor of two for angular separations of 
$\lesssim~8^{\circ}$.  \citet{lin00} and \citet{red00} provide estimates of 
the distance to the edge of the LIC.  With these estimates, the angular 
separations of $\lesssim~8^{\circ}$ correspond to physical distances of 
$\lesssim~0.6$~pc or $\lesssim~10^5$~AU at the edge of the LIC. 
\end{enumerate}

\acknowledgements
The authors would like to thank B. Wood for useful discussions and insightful 
comments regarding this work.  We would also like to thank C. Gry for sending 
the results of O. Dupin's analysis of the $\zeta$~CMa line of sight.  We thank
 the anonymous referee for their careful reading and helpful comments.  This 
research is supported by NASA grants NGT5-50242 and S-56500-D to the 
University of Colorado at Boulder.

\clearpage
\begin{center}
\begin{deluxetable}{ccccc}
\label{table1}
\tabletypesize{\small}
\tablewidth{0pt}
\tablecaption{Summary of STIS Observations}
\tablehead{ HD & Exposure & Date & S/N & S/N\\
\# & Time && Mg~II k & Mg~II h\\ 
& (s) && (2796.352 \AA) & (2803.531 \AA)\\}
\startdata
21847 & 180      & 1999 Jan 26     & 10      & \phn 7\\
26345 & 144      & 1999 Jan \phn 9 & \phn 9  & \phn 8\\
26784 & 288      & 1999 Aug \phn 3 & 18      & 14    \\
27561 & 120      & 1998 Nov \phn 7 & 11      & \phn 9\\
27808 & 312      & 1999 Feb \phn 2 & 19      & 15    \\
27848 & 144      & 1999 Feb 10     & \phn 8  & \phn 7\\
28033 & 288      & 1999 Jul 31     & \phn 6  & \phn 5\\
28205 & 240      & 1999 Feb \phn 4 & 15      & 11    \\
28237 & 1695\phn & 1998 Dec 23     & 16      & 11    \\
28406 & 240      & 1998 Nov 20     & 10      & \phn 9\\
28483 & 180      & 1998 Nov \phn 8 & 11      & \phn 9\\
28568 & 144      & 1998 Nov \phn 8 & \phn 8  & \phn 7\\
28608 & 240      & 1998 Nov \phn 7 & 13      & 10    \\
28736 & 144      & 1999 Jan 14     & \phn 9  & \phn 7\\
29225 & 240      & 1998 Oct 26     & 12      & 10    \\
29419 & 288      & 1998 Nov \phn 5 & 18      & 14    \\
30738 & 240      & 1998 Nov \phn 7 & 12      & \phn 9\\
31845 & 150      & 1998 Nov \phn 8 & \phn 9  & \phn 7\\
\enddata
\end{deluxetable}
\end{center}

\clearpage
\begin{center}
\begin{deluxetable}{ccccccccc}
\label{table2}
\tabletypesize{\small}
\tablewidth{0pt}
\tablecaption{Properties of Hyades Stars}
\tablehead{ HD & Spectral & $m_{\rm V}$ & $v_{\rm R}$\tablenotemark{a}  & 
$l$ & $b$ & distance \tablenotemark{b} & 
$v_{\rm{LIC}}$\tablenotemark{c} & $\log N_{\rm{HI}}$(LIC)\tablenotemark{d} \\
\# & Type && (km s$^{-1}$) & ($^{\circ}$) & ($^{\circ}$) & (pc) & 
(km s$^{-1}$)  & $\log (\rm cm^{-2})$}
\startdata
21847 & F8\phantom{V}  & 7.3 &       & 156.2 & --16.6 & 48.9 & 22.6 & 18.31 \\
26345 & F6V            & 6.6 & +37.1 & 175.2 & --23.6 & 43.1 & 25.1 & 18.23 \\
26784 & F8V            & 7.1 & +38.5 & 182.4 & --27.9 & 47.4 & 25.1 & 18.05 \\
27561 & F5V            & 6.6 & +39.2 & 180.4 & --24.3 & 51.4 & 25.3 & 18.14 \\
27808 & F8V            & 7.1 & +38.9 & 174.8 & --19.1 & 40.9 & 25.2 & 18.23 \\
27848 & F6V            & 7.0 & +40.1 & 178.6 & --22.0 & 53.4 & 25.4 & 18.18 \\
28033 & F8V            & 7.4 & +38.8 & 175.4 & --18.9 & 46.4 & 25.3 & 18.23 \\
28205 & F8V            & 7.4 & +39.3 & 180.4 & --22.4 & 45.8 & 25.4 & 18.15 \\
28237 & F8\phantom{V}  & 7.5 & +40.2 & 183.7 & --24.7 & 47.2 & 25.4 & 18.07 \\
28406 & F6V            & 6.9 & +38.6 & 178.8 & --20.6 & 46.3 & 25.4 & 18.18 \\
28483 & F6V            & 7.1 & +38.0 & 177.3 & --19.2 & 50.2 & 25.4 & 18.21 \\
28568 & F5V            & 6.5 & +40.9 & 180.5 & --21.4 & 41.2 & 25.5 & 18.15 \\
28608 & F5\phantom{V}  & 7.0 & +41.4 & 185.1 & --24.7 & 43.6 & 25.4 & 18.06 \\
28736 & F5V            & 6.4 & +39.8 & 190.2 & --27.6 & 43.2 & 25.2 & 18.00 \\
29225 & F5V            & 6.6 & +33.7 & 181.6 & --20.5 & 43.5 & 25.6 & 18.14 \\
29419 & F5\phantom{V}  & 7.5 & +39.9 & 176.0 & --15.6 & 44.2 & 25.3 & 18.22 \\
30738 & F8\phantom{V}  & 7.3 & +42.7 & 183.5 & --17.6 & 51.8 & 25.7 & 18.13 \\
31845 & F5V            & 6.8 & +42.5 & 185.1 & --16.0 & 43.3 & 25.7 & 18.12 \\
\enddata
\tablenotetext{a}{Taken from \citet{perry98}, see references within.}
\tablenotetext{b}{$Hipparcos$ distances \citep{perry97}.}
\tablenotetext{c}{Predicted projected LIC velocity from \citet{lall92}.}
\tablenotetext{d}{Predicted hydrogen column density from \citet{red00}.}
\end{deluxetable}
\end{center}

\clearpage
\begin{deluxetable}{ccccccccccc}
\label{table3}
\tablewidth{0pt}
\tabletypesize{\tiny}
\tablecaption{Fit Parameters for the Mg II Interstellar Components}
\tablehead{ & \multicolumn{3}{c}{LIC Component} & \multicolumn{3}{c}{Hyades 
Cloud Component} & \multicolumn{3}{c}{Third Component} \\
HD & \colhead{$v$} & \colhead{$b$} & \colhead{$\log N_{\rm MgII}$} & 
\colhead{$v$} & \colhead{$b$} & \colhead{$\log N_{\rm MgII}$} & \colhead{$v$} 
& \colhead{$b$} & \colhead{$\log N_{\rm MgII}$} & $\chi_{\nu}^2$ \\
\# & \colhead{(km s$^{-1}$)}&\colhead{(km s$^{-1}$)}&\colhead{$\log 
(\rm cm^{-2})$} & \colhead{(km s$^{-1}$)}&\colhead{(km s$^{-1}$)} & 
\colhead{$\log (\rm cm^{-2})$ }& \colhead{(km s$^{-1}$)}& 
\colhead{(km s$^{-1}$)}&\colhead{$\log (\rm cm^{-2})$} }
\startdata
21847 & 21.1 $\pm$ 0.3 & 2.7 $\pm$ 1.0 & 13.14 $\pm$ 0.37 &&&&&&& 1.16 \\
26345 & 21.1 $\pm$ 0.5 & 3.8 $\pm$ 0.7 & 12.65 $\pm$ 0.08 & 13.6 $\pm$ 1.1 & 
1.9 $\pm$ 1.3 & 11.94 $\pm$ 0.29 &&&& 0.79 \\
26784 & 23.0 $\pm$ 1.0 & 2.9 $\pm$ 0.6 & 12.32 $\pm$ 0.13 & 15.5 $\pm$ 1.1 & 
3.7 $\pm$ 1.2 & 12.34 $\pm$ 0.12 &&&& 1.02 \\
27561 & 22.2 $\pm$ 0.5 & 4.1 $\pm$ 0.7 & 12.50 $\pm$ 0.06 & 14.4 $\pm$ 0.8 & 
2.5 $\pm$ 1.0 & 12.04 $\pm$ 0.19 &&&& 0.98 \\
27808 & 23.1 $\pm$ 0.4 & 3.4 $\pm$ 0.4 & 12.86 $\pm$ 0.11 &&&&&&& 0.95 \\
27848 & 22.4 $\pm$ 1.0 & 3.8 $\pm$ 1.2 & 12.56 $\pm$ 0.24 & 16.4 $\pm$ 2.2 & 
2.7 $\pm$ 1.6 & 11.88 $\pm$ 0.31 &&&& 0.82 \\
28033 & 23.6 $\pm$ 0.5 & 3.2 $\pm$ 0.8 & 12.89 $\pm$ 0.35 &&&&&&& 0.76 \\
28205 & 23.3 $\pm$ 0.3 & 2.3 $\pm$ 0.6 & 12.59 $\pm$ 0.28 & 14.8 $\pm$ 1.7 & 
4.2 $\pm$ 1.6 & 12.06 $\pm$ 0.14 &&&& 0.82 \\
28237 & 22.4 $\pm$ 0.7 & 3.9 $\pm$ 0.9 & 12.42 $\pm$ 0.12 & 15.6 $\pm$ 1.9 & 
4.0 $\pm$ 1.9 & 12.17 $\pm$ 0.20 &&&& 1.08 \\
28406 & 22.1 $\pm$ 0.3 & 4.3 $\pm$ 0.3 & 12.67 $\pm$ 0.07 &&&&&&& 0.92 \\
28483 & 20.8 $\pm$ 0.3 & 3.8 $\pm$ 0.5 & 12.67 $\pm$ 0.15 &&&&&&& 0.85 \\
28568 & 23.9 $\pm$ 0.5 & 3.3 $\pm$ 0.7 & 12.59 $\pm$ 0.15 & 16.5 $\pm$ 1.6 & 
2.5 $\pm$ 1.6 & 11.93 $\pm$ 0.27 &&&& 1.01 \\
28608 & 23.2 $\pm$ 0.4 & 3.1\tablenotemark{a} $\pm$ 0.4 & 12.42 $\pm$ 0.06 & 
15.8 $\pm$ 0.9 & 3.1\tablenotemark{a} $\pm$ 0.4 & 12.18 $\pm$ 0.08 &&&& 1.06\\
28736 & 21.6 $\pm$ 0.5 & 4.4 $\pm$ 0.6 & 12.65 $\pm$ 0.05 & 13.4 $\pm$ 0.7 & 
1.7 $\pm$ 0.9 & 12.06 $\pm$ 0.25 & --4.3 $\pm$ 0.4 & 2.4 $\pm$ 0.6 & 12.25 
$\pm$ 0.15 & 0.83 \\
29225 & 22.5 $\pm$ 0.2 & 3.7 $\pm$ 0.3 & 12.68 $\pm$ 0.04 &&&&&&& 0.96 \\
29419 & 23.2 $\pm$ 0.2 & 3.1 $\pm$ 0.3 & 12.69 $\pm$ 0.08 & 12.3 $\pm$ 0.2 & 
2.1 $\pm$ 1.2 & 12.66 $\pm$ 0.33 &&&& 1.00 \\
30738 & 20.3 $\pm$ 0.4 & 3.8 $\pm$ 0.5 & 12.48 $\pm$ 0.10 &&&&&&& 0.98 \\
31845 & 22.4 $\pm$ 0.3 & 3.7 $\pm$ 0.5 & 12.49 $\pm$ 0.05 &&&&&&& 1.00 \\
\enddata
\tablenotetext{a}{The low signal-to-noise and blended absorption of the Mg~II 
lines observed along the line of sight to HD~28608 did not allow for a unique 
solution.  Therefore, based on the line analysis of the nearby star HD~28237, 
the Doppler parameter for both components were forced to be equal.}
\end{deluxetable}  

\clearpage
\begin{center}
\begin{deluxetable}{cccccc}
\label{table4}
\tabletypesize{\small}
\tablewidth{0pt}
\tablecaption{Comparison of Measured and Predicted LIC Parameters for Each 
Line of Sight}
\tablehead{ HD & $v_{\rm{LIC}}$\tablenotemark{a} & 
$v_{\rm{LIC}}$\tablenotemark{b} & 
$\log N_{\rm{HI}}$(LIC)\tablenotemark{c} & 
$\log N_{\rm{MgII}}$(LIC)\tablenotemark{d} & 
D(Mg) \\
\# & (km s$^{-1}$)  & (km s$^{-1}$)  & $\log (\rm cm^{-2})$ & 
$\log (\rm cm^{-2})$ &}
\startdata
21847 & 22.6 $\pm$ 0.5 & 21.1 $\pm$ 0.3 & 18.31 & 13.14 $\pm$ 0.37 & --0.76 
$\pm$ 0.37\\
26345 & 25.1 $\pm$ 0.5 & 21.1 $\pm$ 0.5 & 18.23 & 12.65 $\pm$ 0.08 & --1.17 
$\pm$ 0.08 \\
26784 & 25.1 $\pm$ 0.5 & 23.0 $\pm$ 1.0 & 18.05 & 12.32 $\pm$ 0.13 & --1.32 
$\pm$ 0.13\\
27561 & 25.3 $\pm$ 0.5 & 22.2 $\pm$ 0.5 & 18.14 & 12.50 $\pm$ 0.06 & --1.23 
$\pm$ 0.06 \\
27808 & 25.2 $\pm$ 0.5 & 23.1 $\pm$ 0.4 & 18.23 & 12.86 $\pm$ 0.11 & --0.96 
$\pm$ 0.11 \\
27848 & 25.4 $\pm$ 0.5 & 22.4 $\pm$ 1.0 & 18.18 & 12.56 $\pm$ 0.24 & --1.21 
$\pm$ 0.24 \\
28033 & 25.3 $\pm$ 0.5 & 23.6 $\pm$ 0.5 & 18.23 & 12.89 $\pm$ 0.35 & --0.93 
$\pm$ 0.35 \\
28205 & 25.4 $\pm$ 0.5 & 23.3 $\pm$ 0.3 & 18.15 & 12.59 $\pm$ 0.28 & --1.15 
$\pm$ 0.28 \\ 
28237 & 25.4 $\pm$ 0.5 & 22.4 $\pm$ 0.7 & 18.07 & 12.42 $\pm$ 0.12 & --1.24 
$\pm$ 0.12 \\
28406 & 25.4 $\pm$ 0.5 & 22.1 $\pm$ 0.3 & 18.18 & 12.67 $\pm$ 0.07 & --1.10 
$\pm$ 0.07 \\
28483 & 25.4 $\pm$ 0.5 & 20.8 $\pm$ 0.3 & 18.21 & 12.67 $\pm$ 0.15 & --1.13 
$\pm$ 0.15 \\
28568 & 25.5 $\pm$ 0.5 & 23.9 $\pm$ 0.5 & 18.15 & 12.59 $\pm$ 0.15 & --1.15 
$\pm$ 0.15 \\
28608 & 25.4 $\pm$ 0.5 & 23.2 $\pm$ 0.4 & 18.06 & 12.42 $\pm$ 0.06 & --1.23 
$\pm$ 0.06 \\
28736 & 25.2 $\pm$ 0.5 & 21.6 $\pm$ 0.5 & 18.00 & 12.65 $\pm$ 0.05 & --0.94 
$\pm$ 0.05 \\
29225 & 25.6 $\pm$ 0.5 & 22.5 $\pm$ 0.2 & 18.14 & 12.68 $\pm$ 0.04 & --1.05 
$\pm$ 0.04 \\
29419 & 25.3 $\pm$ 0.5 & 23.2 $\pm$ 0.2 & 18.22 & 12.69 $\pm$ 0.08 & --1.12 
$\pm$ 0.08 \\
30738 & 25.7 $\pm$ 0.5 & 20.3 $\pm$ 0.4 & 18.13 & 12.48 $\pm$ 0.10 & --1.24 
$\pm$ 0.10 \\
31845 & 25.7 $\pm$ 0.5 & 22.4 $\pm$ 0.3 & 18.12 & 12.49 $\pm$ 0.05 & --1.22 
$\pm$ 0.05 \\
\enddata
\tablenotetext{a}{Predicted projected LIC velocity from \citet{lall92}.}
\tablenotetext{b}{Observed projected LIC velocity, see Table 3.}
\tablenotetext{c}{Predicted hydrogen column density from \citet{red00}.}
\tablenotetext{d}{Observed Mg~II column density, see Table 3.}
\end{deluxetable}
\end{center}

\clearpage
\begin{center}
\begin{deluxetable}{llcccccccccl}
\label{table5}
\tabletypesize{\tiny}
\tablewidth{0pt}
\tablecaption{HST Targets in the Vicinity of the Hyades Stars}
\tablehead{ HD & Other & $l$ & $b$ & distance \tablenotemark{a} & $v$ & 
$b_{\rm MgII}$ & $\log N_{\rm MgII}$ & Temperature &$\xi$& D(Mg)& Reference \\
\# & Name & ($^{\circ}$) & ($^{\circ}$) & (pc) & (km s$^{-1}$) & 
(km s$^{-1}$)  & $\log (\rm cm^{-2})$ & (K) &(km s$^{-1}$)&&}
\startdata
& \multicolumn{5}{c}{Targets with absorption at the Hyades Cloud velocity:} 
&&&&&& \\
\cline{2-6} 

48915 & Sirius       & 227.2 & \phn --8.9  &  \phn\phn 2.6 & $13.1\phn \pm 
1.5\phn$ & $3.1\phn \pm 0.4\phn$ & 12.00 $\pm$ 0.05 & $3000^{+2000}_{-1000}$ 
& $2.7\phn \pm 0.3$ & $-0.99 \pm 0.21$ & 1,2 \\
20630 & $\kappa^1$ Cet&178.2 & --43.1 &  \phn\phn 9.2 & $13.32 \pm 0.05$ & 
$2.64 \pm 0.10$ & 12.19 $\pm$ 0.03 & $\lesssim~5000$ & 1.8-2.7 & \nodata & 3\\
11443 & $\alpha$ Tri & 138.6 & --31.4 & \phn 19.7 & $14.2\phn \pm 0.4\phn$ & 
$4.1\phn \pm 1.4\phn$ & 12.76 $\pm$ 0.10 & $8300~\pm~2000$ & $<$ 1.7 & 
$-0.83 \pm 0.14$ & 4 \\

22468 & HR 1099      & 184.9 & --41.6 & \phn 29.0 & $14.8\phn \pm 0.1\phn$ & 
$2.42 \pm 0.08$ & 11.74 $\pm$ 0.02 & \nodata & \nodata & $-1.44 \pm 0.16$ & 
5 \\

44402 & $\zeta$ CMa  & 237.5 & --19.4 &103.4 & $13.0\phn \pm 1.5\phn$ & 
\nodata & \nodata & 6400-8400 & 1.8-2.2 & \nodata & 6 \\

52089 & $\epsilon$ CMa&239.8 & --11.3 &132.1 & $10\phantom{.0} \pm 2\phn\phn$ 
& $2.38 \pm 0.04$ &     12.00 $\pm$ 0.05 & 3600 $\pm$ 1500 & $1.85 \pm 0.3$ & 
$-0.47 \phantom{\pm 1.111}$ & 7, 8 \\

& \multicolumn{5}{c}{Targets without absorption at the Hyades Cloud 
velocity:} &&&&&& \\
\cline{2-6} 
22049 & $\epsilon$ Eri & 195.8 & --48.1 &  \phn\phn 3.2 & & & & & & & 4\\
26965 & 40 Eri A       & 200.8 & --38.1 &  \phn\phn 5.0 & & & & & & & 9 \\
39587 & $\chi^1$ Ori   & 188.5 & \phn --2.7&\phn\phn 8.7 & & & & & & & 3 \\
34029 & Capella        & 162.6 & \phantom{--1}4.6 & \phn 12.9 &&&&&&& 10, 11\\
\phn\phn 432&$\beta$ Cas&117.5 & \phn --3.3  & \phn 16.7 & & & & & & & 4 \\
\phn 1405&PW And       & 114.6 & --31.4 & \phn 21.9 & & & & & & & 12 \\
\phn 8538&$\delta$ Cas & 127.2 & \phn --2.4  & \phn 30.5 & & & & & & & 13 \\
      & G191-B2B       & 156.0 & \phantom{--1}7.1 & \phn 68.8&&&&&&& 14 \\
\enddata
\tablenotetext{a}{$Hipparcos$ distances \citep{perry97}.}
\tablerefs{(1) Lallement et al. 1994; (2) H\'{e}brard et al. 1999; (3) 
Redfield et al. in preparation; (4) Dring et al. 1997; (5) Piskunov et al. 
1997; (6) Dupin 1998; (7) Gry et al. 1995; (8) Gry \& Jenkins in press; (9) 
Wood \& Linsky 1998; (10) Linsky et al. 1993; (11) Linsky et al. 1995; (12) 
Wood et al. 2000; (13) Lallement \& Ferlet 1997; (14) Sahu et al. 1999;}
\end{deluxetable}
\end{center}

\clearpage
\begin{center}
\begin{deluxetable}{llccccccl}
\label{table6}
\tabletypesize{\small}
\tablewidth{0pt}
\tablecaption{Ca~II Targets in the Vicinity of the Hyades Stars}
\tablehead{ HD & Other & $l$ & $b$ & distance \tablenotemark{a} & 
$v$ & $b_{\rm CaII}$ & $\log N_{\rm CaII}$ & Reference \\
\# & Name & ($^{\circ}$) & ($^{\circ}$) & (pc) & 
(km s$^{-1}$)  & (km s$^{-1}$)  & $\log (\rm cm^{-2})$ &}
\startdata
& \multicolumn{5}{c}{Targets with absorption at the  Hyades Cloud velocity:} 
&&& \\
\cline{2-6}
32630 & $\eta$ Aur   & 165.4 & \phantom{--1}0.3    & \phn 67.2  & 10.7 $\pm$ 
0.1 & 1.70 $\pm$ 0.10 & 10.75 $\pm$ 0.01 & 1 \\
                                          &&&&& 10.7 $\pm$ 0.3 & 1.51 $\pm$ 
0.3\phn & 10.83 $\pm$ 0.02 & 2 \\
35468 & $\gamma$ Ori & 196.9 & --16.0 & \phn 74.5  & 16.0 $\pm$ 0.1 & 1.90 
$\pm$ 0.20 & 10.34 $\pm$ 0.04 & 1 \\
23630 & $\eta$ Tau   & 166.7 & --23.5 & 112.7 & 15.1 $\pm$ 0.1 & 2.10 $\pm$ 
0.18 & 10.56 $\pm$ 0.03 & 1 \\
                         &&&&& 15.2 $\pm$ 0.3 & 1.30\tablenotemark{b} $\pm$ 
0.3\phn  & 10.70 $\pm$ 0.03 & 2 \\
23302 & 17 Tau       & 166.2 & --23.9 & 113.6 & 15.7 $\pm$ 0.3 & 1.32 $\pm$ 
0.3\phn  & 10.60 $\pm$ 0.03 & 2 \\
22928 & $\delta$ Per & 150.3 & \phn --5.8  & 161.8 & 16.2 $\pm$ 0.1 & 1.60 
$\pm$ 0.19 & 10.48 $\pm$ 0.02 & 1 \\
& \multicolumn{5}{c}{Targets without absorption at the Hyades Cloud 
velocity:} &&& \\
\cline{2-6}
16970 & $\gamma$ Cet & 168.9 & --49.4 & \phn 25.1 & & & & 3 \\
19356 & $\beta$ Per  & 149.0 & --14.9 & \phn 28.5 & & & & 1 \\ 
35497 & $\beta$ Tau  & 178.0 & \phn --3.8  & \phn 40.2 & & & & 1 \\
25642 & $\lambda$ Per& 151.6 & \phn --1.3  & 106.3 & & & & 1 \\ 
\phn 3360  & $\zeta$ Cas  & 120.8 & \phn --8.9  & 183.2 & & & & 1 \\
\phn 5394  & $\gamma$ Cas & 123.6 & \phn --2.2  & 188.0 & & & & 1, 2 \\
\phn 3369  & $\pi$ And    & 119.5 & --29.1 & 201.2 & & & & 1 \\
\phn 4727  & $\nu$ And    & 122.6 & --21.8 & 208.3 & & & & 1 \\
\enddata
\tablenotetext{a}{$Hipparcos$ distances \citep{perry97}.}
\tablenotetext{b}{fixed b-value \citep{welty96}.}
\tablerefs{(1) Vallerga et al. 1993; (2) Welty et al. 1996; (3) Crawford et 
al. 1998;}
\end{deluxetable}
\end{center}

\clearpage
Fig. 8 - Difference in the Mg~II column density between pairs of stars as a 
function of angular separation (degrees).  The top panel displays the 
difference in LIC Mg~II column densities (open circles) and 1$\sigma$ error 
bars for all pairings of the 18 total targets.  The cross-hatch symbols 
($\times$) indicate the difference in the predicted LIC H~I column densities, 
provided by the \citet{red00} LIC model.  The long dashed line indicates zero 
change, and the short dashed line indicates a factor of 2 difference in column
 density.  We expect the difference in column density to increase with 
increasing angular distance.  A factor of 2 difference is consistently 
obtained for objects separated by $\gtrsim~8^{\circ}$.  Because we have an 
indication of the distance to the edge of the LIC \citep{red00}, we can 
estimate the physical distance at the edge of the LIC between the lines of 
sight to a pair of stars from their angular separation.  The top axis of the 
LIC plot shows physical distance in units of parsecs.  The bottom panel 
displays the difference in the Hyades Cloud Mg~II column density and 1$\sigma$
 error bars for all pairings of the 10 targets that show Hyades Cloud 
absorption.  The format of the bottom panel is identical to the top.  We 
expect differences in column density to be larger at smaller angular 
distances, since the Hyades Cloud is at a greater distance than the LIC, 
meaning the angular distances correspond to larger physical distances.
\clearpage

\clearpage
Fig. 9 - Difference in the Mg~II Doppler parameter between pairs of stars as 
a function of angular separation (degrees).  The top panel displays the 
difference in LIC Mg~II Doppler parameters and 1$\sigma$ error bars for all 
pairings of the 18 total targets.  The long dashed line indicates zero change.
  Because we have an indication of the distance to the edge of the LIC 
\citep{red00}, we can estimate the physical distance at the edge of the LIC 
between the lines of sight to a pair of stars from their angular separation.  
The top axis of the LIC plot shows physical distance in units of parsecs.  The
 bottom panel displays the difference in the Hyades Cloud Mg~II column density
 and 1$\sigma$ error bars for all pairings of the 10 targets that show Hyades 
Cloud absorption.  The format of the bottom panel is identical to the top.
\clearpage

\end{document}